\documentclass[journal]{IEEEtran}
\usepackage{url}
\usepackage{cite}
\usepackage{soul,color}
\usepackage{xcolor}
\usepackage{array}
\usepackage{enumitem}
\usepackage{multirow}
\usepackage{times}
\usepackage{graphicx,amsmath,amssymb}
\usepackage[ruled,vlined]{algorithm2e}
\usepackage{algorithmic}
\usepackage{setspace}
\DeclareMathSizes{10}{8}{7}{6}
\IEEEaftertitletext{\vspace{-2\baselineskip}}
\include{pythonlisting}
\newtheorem{lemma}{Lemma}

\begin{document}
\bstctlcite{IEEEexample:BSTcontrol}
\title{Performance Analysis and Optimization of NOMA with HARQ for Short Packet Communications in Massive IoT}

\author{Fatemeh~Ghanami,
        Ghosheh Abed Hodtani, 
        Branka~Vucetic,
        Mahyar~Shirvanimoghaddam
\thanks{F. Ghanami and G. A. Hodtani are with Ferdowsi University of Mashhad, Mashhad, Iran  (e-mail:
f.ghanami@mail.um.ac.ir, ghodtani@gmail.com).}
\thanks{F. Ghanami, B. Vucetic, and M. Shirvanimoghaddam are with the Centre of IoT and Telecommunications, The University of Sydney, Sydney, NSW 2006, Australia (e-mail:\{fatemeh.ghanami,mahyar.shirvanimoghaddam, branka.vucetic\}@sydney.edu.au).}
\thanks{Copyright (c) 2020 IEEE. Personal use of this material is permitted. However, permission to use this material for any other purposes must be obtained from the IEEE by sending a request to pubs-permissions@ieee.org.}}

\maketitle
\begin{abstract}

In this paper, we consider the massive non-orthogonal multiple access (NOMA) with hybrid automatic repeat request (HARQ) for short packet communications. To reduce the latency, each user can perform one re-transmission provided that the previous packet was not decoded successfully. The system performance is evaluated for both coordinated and uncoordinated transmissions.  We first develop a Markov model (MM) to analyze the system dynamics and characterize the packet error rate (PER) and throughput of each user in the coordinated scenario.  The power levels are then optimized for two scenarios, including the power constrained and reliability constrained scenarios. A simple yet efficient dynamic cell planning is also designed for the uncoordinated scenario. Numerical results show that both coordinated and uncoordinated NOMA-HARQ with a limited number of retransmissions can achieve the desired level of reliability with the guaranteed latency using a proper power control strategy. Results also show that NOMA-HARQ achieves a higher throughput compared to the orthogonal multiple access scheme with HARQ under the same average received power constraint at the base station. 
\end{abstract}
\begin{IEEEkeywords}
Finite block length, hybrid automatic repeat request (HARQ), non-orthogonal multiple access (NOMA).
\end{IEEEkeywords}
\IEEEpeerreviewmaketitle

\vspace{-2ex}
\section{Introduction}
\vspace{-1ex}
\IEEEPARstart{W}{ireless} cellular communications has been historically concentrated on human-centric communications to mainly increase the spectral efficiency \cite{cirik2019toward}. The fifth generation (5G) of mobile standards expands its focus into three usage scenarios, including enhanced mobile broadband (eMBB), massive machine Type communication (mMTC) tailored for many internet of things (IoT) applications, and ultra-reliable low-latency communications (URLLC) for mission-critical applications \cite{shafi20175g}. URLLC poses two conflicting performance requirements, low latency and ultra-high reliability. That is a target packet error rate (PER) of less than $10^{-5}$ should be reached within a user plane latency of 1ms \cite{chen2018ultra}. On the other hand, mMTC requirements are massive connectivity, high energy efficiency, and low cost.  In mMTC, the system design should be flexible to provide various requirements of latency and reliability \cite{durisi2016toward} for a large number of devices, which is expected to increase to 50 billion by 2030 \cite{statista}. To address these challenging requirements, 5G New Radio (NR) explores several novel access mechanism, such as  non-orthogonal multiple access (NOMA) \cite{cirik2019toward}.

\subsection{Related Works}
Due to massive connectivity and demand growth for various IoT services, the implementation of 5G systems necessitates more effective radio access technologies. NOMA has recently emerged as a solution to improve the spectral efficiency compared to orthogonal multiple access (OMA) and showed significant improvements in terms of throughput and system capacity in both mMTC and URLLC \cite{shirvanimoghaddam2017massive,ding2017application}. In NOMA, multiple users share the same resources, e.g., time, frequency, and space, by applying power-domain multiplexing or code-domain multiplexing \cite{dai2015non,yuan2016non}. Compared to OMA, NOMA was shown to exploit the channel diversity more effectively through successive interference cancellation (SIC) \cite{xu2017energy,yu2017performance}. NOMA has been mainly designed \cite{ding2017application, mlika2020massive,dai2015non,shirvanimoghaddam2017massive,dinh2020low,shirvanimoghaddam2017fundamental,liang2019non,liu2019stable} based on the classical Shannon capacity formula, which is accurate when the block length is asymptotically long. These works provide useful insights on the effectiveness of NOMA for many potential applications and its advantages in terms of throughput and scalability. However, there is not much known about the performance of NOMA for short packet communications, which is the scenario of interest for many mMTC and URLLC applications. In the finite block-length regime, due to finite number of channel observations, the coding gain is reduced and the gap to the Shannon’s limit is increased \cite{polyanskiy2010channel,shirvanimoghaddam2018short}. Short packet communication is necessary for URLLC and mMTC to minimize latency when the available resources are limited \cite{bennis2018ultrareliable,shirvanimoghaddam2018short}. Recently, NOMA with short packet communications has been investigated  for both uplink (e.g. \cite{8957374, 9097306, 9149137, 8917650}) and downlink (e.g. \cite{8772114, 9013299, 9013628, 9141422, 9096300}) scenarios.

In \cite{amjad2019noma,amjad2020effective,yu2017performance,sun2018short,dosti2019performance,choi2020opportunistic, xiang2020noma}, authors studied NOMA in the finite block length regime and compared it with OMA in terms of latency and reliability. The results of these findings make the foundation for potential applications of NOMA in URLLC and mMTC scenarios. Re-transmission techniques were also considered in conjunction with NOMA in the finite block length regime \cite{hu2020throughput}. In particular, mMTC benefits from retrasnmissions as it provides time diversity to increase reliability and accordingly the coverage. This is mainly because many mMTC applications has to deal with sporadic traffic of small payloads from each device; therefore, a packet can be retransmitted  several times to be successfully delivered at the base station \cite{shirvanimoghaddam2017massive}. The concept was mainly adopted in the long-term evolution (LTE) release 13 to 15, e.g. LTE-M and NB-IoT \cite{GSMA-LTE}, to address the fast-expanding market for low power wide area connectivity. However, the access technology is still orthogonal, where the devices need to be identified and allocated orthogonal resources for their transmissions. On the other hand in URLLC, the retransmission will improve reliability when other diversity resources are not available. However, due to critical latency requirements the number of retransmissions must be kept low \cite{bennis2018ultrareliable,shirvanimoghaddam2018short}.

Wireless networks usually adopt Automatic Repeat reQuest (ARQ) and its variants, including Hybrid ARQ (HARQ), for retransmissions when required.  A feedback message is sent from the receiver to inform the transmitter whether a retransmission is required or not.  In Chase combining (CC) HARQ, the entire codeword is sent in each retransmission and repeated packets are combined using the maximum ratio combining (MRC) approach to  increase the effective signal to noise ratio (SNR). In incremental redundancy (IR) HARQ, the original codeword is divided into multiple sub-codewords, which are sent in subsequent retransmissions to increases the coding gain. Both sub-classes are being actively investigated in the finite block length regime  \cite{makki2014finite,makki2018fast}. HARQ was recently considered for downlink NOMA with 2 users \cite{xu2019harq,cai2018performanc,cai2018outage,xu2018average}, where the power allocation, rate selection, and outage performance were analyzed in the infinite block length regime. Authors in \cite{kotaba2019improving} proposed a NOMA based retransmission strategy for uplink NOMA, where the freshly generated packets can share the same radio resources with the retransmitted packets. Results show that a significant latency reduction can be achieved, which makes it a suitable retransmission technique for URLLC. In \cite{chandran2019novel}, a novel retransmission scheme was proposed for two-user downlink NOMA employing HARQ, where the power level of users are adjusted in the retransmission to potentially reduce the number of attempts. Results showed that the proposed approach can increase the cell throughput. The analysis and design are based on the infinitely long block length assumption.

The two-user donwlink NOMA with HARQ under the finite block length assumption was also considered in the literature \cite{yu2017performance,huang2019block,marasinghe2019block,xu2017energy}, where the throughput and packet error rate were analyzed and power level optimization was carried out. While these papers shed light on the overall performance of NOMA with retransmissions in the finite block length regime, NOMA with HARQ for short packet communications in the uplink of cellular systems has not been thoroughly investigated. 
\vspace{-2ex}
\subsection{Contributions}
In this paper, we take steps towards understanding the performance of multi-user uplink NOMA with HARQ in the finite block length regime and shed light on its application in massive IoT with guaranteed delay and reliability performance. The main contributions of the paper are as follows. 
\begin{itemize}[leftmargin=*]
\item We propose a Markov model to understand the dynamics of the uplink NOMA-HARQ with one retransmission for an arbitrary number of active users. We characterize the SIC and its decoding order and accordingly the state transition probabilities. 
\item Using the proposed Markov model, the packet error rate, packet delivery delay, and throughput for each user are analyzed. Two different optimization problems are defined and solved numerically for the power constrained and reliability constrained scenarios. We show that using NOMA-HARQ with a limited number of retransmissions, the desired level of reliability and throughput can be maintained for all active users. 
\item We propose a dynamic cell planning scheme for the uncoordinated scenario, where the users that are randomly located in the cell adjust their power levels according to cell planning information sent by the base station (BS). We show that the proposed cell planning scheme can effectively accommodate a large number of active users in an uncoordinated manner. 
\item We finally shed light on the application of the NOMA-HARQ for grant-free access \cite{abbas2020grant,liu2020analyzing}, where each user randomly chooses a subband and a power level according to its location, and transmits its message using HARQ with one retransmission. Although the number of active users in each subband is random, we show that using the proposed cell planning with a proper number of cell segments and power levels, the desired level of reliability and throughput can be achieved for all users. 
\end{itemize}
\vspace{-2ex}
\subsection{Paper Organization}
The rest of the paper is organized as follows. Section II presents the system model and some preliminaries on NOMA and finite block length analysis. In Section III, we propose a Markov model to analyze the performance of the uplink NOMA-HARQ scheme and characterize the packet error rate and throughput.  The optimization of the power splitting ratios are presented in Section IV. Then, we present the proposed dynamic cell planning for the uncoordinated transmission using NOMA-HARQ in Section V. Numerical results are presented in Section VI followed by concluding remarks in Section VII. 
\vspace{-2ex}
\section{System Model and Preliminaries}
\subsection{Channel Model}
We consider an uplink non-orthogonal multiple access cellular system employing HARQ with a limited number of re-transmissions. It is assumed that users and the BS are equipped with single antenna each. The BS is located at the origin of the cell and devices are randomly deployed in the cell in fixed locations.

Similar to \cite{zeng2019energy}, the channel between the $i^{th}$ user and the BS, denoted by $g_i$, is characterized by small-scale Rayleigh fading and large scale path-loss. We assume block fading under which the channel gain is constant over a time block and varies independently between the blocks. In particular, the received power, $P_i$, from the $i^{th}$ user located at distance $d_i$ from the BS is given by:
\begin{equation}
    P_i = |g_i|^2 P_{\mathrm{t},i},
\end{equation}
where $P_{\mathrm{t},i}$ is the transmit power of user $i$, $|g_i|^2 = h_i d_i^{-\rho}$, $h_i$ is the small-scale fading with exponential distribution, i.e., $h_i\sim\exp{(1)}$, and $\rho$ is the path loss exponent. A summary of notations commonly used in the paper is listed in Table \ref{table:1}.
\vspace{-2ex}
\subsection{NOMA transmission with HARQ}
Unlike the orthogonal multiple access, where BS allocates to each user a designated radio resource depending on its priority to transmit its packets, NOMA allows multiple users to share the same radio resources. In particular, the received signal at the BS at time instance $t$, denoted by $y(t)$, is represented by
\begin{table}[t]
\centering
\caption{List of Notations.}
\label{table:1}
\scriptsize
\begin{tabular}{ |c|p{6cm}|}
\hline
{\textbf{Notation}} & \textbf{Description}\\
\hline
$e_t$ & The target packet error rate\\
\hline
$\eta_i$ & Throughput of the $i^{th}$ user\\
\hline
$e_i$ & PER of the $i^{th}$ user\\
\hline
$d_i$& The distance between the $i^{th}$ user and the BS\\
\hline
$g_i$& The gain of the channel between the $i^{th}$ user and the BS\\
\hline
$P_{\mathrm{t},i}$& The transmit power of the $i^{th}$ user\\
\hline
$k$ & The information block length of each user\\
\hline
$n$ & Codeword length of each user (block length)\\
\hline
$\alpha_i$&The $i^{th}$ power ratio\\
\hline
$\rho$&Path loss exponent\\
\hline
$N$ & The total number of active users\\
\hline
$\mathcal{N}$ & The set of active users\\
\hline
$P_0$ & The total received power at the BS\\
\hline
$R_o$& The outer radius of the cell\\
\hline
$r_i$& The radius of the $i^{th}$ cell ring in the dynamic cell planning\\
\hline
\end{tabular}
\end{table}
\begin{equation}
  y(t)=\sum_{i=1}^N  g_i x_i(t)+w(t),
\end{equation}
where $N$ is the number of active devices who shared the same radio resource via NOMA, $x_i(t)$ is the message transmitted by user $i$, where we assume that $\mathbb{E}[|x_i(t)|^2] = 1$, and $w(t)$ is a circular symmetric white Gaussian noise with unit variance, i.e., $w(t)\sim\mathcal{CN}(0,1)$. 
 
Similar to \cite{choi2017noma}, we assume the channel between the BS and each user is reciprocal, which is considered to be valid in time division duplexing (TDD) systems. The BS can send a beacon signal at the beginning of a time slot to synchronize uplink transmissions, which can be used as a pilot signal to allow each user to estimate its channel to the BS. The users can then perform power control so that their received power at the BS is at a certain level. In fact, the BS allocates the power levels to cell segments. Then according to this power level, the users in that cell segment will adjust their transmit power in such a way that their received power at the BS is at the level of the allocated power. How to allocate power levels to each cell segment by the BS is explained in more details in Section V. In particular, we assume that the $i^{th}$ user performs power control such that its received power at the BS is $P_i=\alpha_i P_0$, where $\alpha_{i}$ is the power splitting ratio and $P_0$ is the desired total received power at the BS. The BS determines the power splitting ratios and broadcasts them to the users. We later show how the power splitting ratios can be determined for coordinated and uncoordinated strategies. 

The BS performs successive interference cancellation (SIC) to decode each active user's message. Without loss of generality, we assume $\alpha_1\ge\alpha_2\ge\cdots\ge\alpha_N$, therefore, the BS first decodes user 1's message by considering other users' signals as noise. If the decoding is successful, user 1's signal will be removed from the received signal and the BS proceeds with the decoding of user 2. This will continues until all users are decoded or the decoding failed for a user. The BS sends an individual instantaneous acknowledgment (ACK) to each user to inform it about the decoding status\footnote{The analysis presented in this paper can be easily extended to consider delayed feedback.}. If the decoding succeeds and the user received an ACK, it will send a new packet in the next time slot. Otherwise, the user will receive a negative acknowledgement (NACK) and it retransmits the previous packet. We consider the Chase combining HARQ (CC-HARQ) where the transmitter sends the exact same packet as the original packet when the retransmission is requested. We assume that the BS performs maximum ratio combing (MRC) to combine received copies of the same packet. It is important to note that due to MRC the decoding order may change. The BS always performs decoding in the descending order of effective signal to interference plus noise ratio (SINR). 
\vspace{-2ex}
\subsection{Packet error rate in the finite block length regime}
We use normal approximation \cite{polyanskiy2010channel} to characterize the packet error rate in the finite block length regime. 
Let $m$ denote the number of copies of a packet at the BS, and $\gamma_i$ denote the SINR of the $i^{th}$ copy of the packet. The packet error rates for CC-HARQ and IR-HARQ in the finite block length regime are then given by \cite{polyanskiy2010channel}:  
\begin{equation}
    \epsilon_{\mathrm{cc}}\left(\gamma_{\mathrm{cc}}\right)    \approx Q\left(\frac{n \log_2\left(1+\gamma_{\mathrm{cc}}\right)-k+\log_2 (n)} {\sqrt{n V\left(\gamma_{\mathrm{cc}}\right)}}\right),
    \label{cc}
\end{equation}
\begin{equation}
   \epsilon_{\mathrm{ir}}\left(\Gamma^{(m)}\right)    \approx Q\left(\frac{n\sum_{i=1}^{m} \log_2(1+\gamma_i)-k+\log_2 (mn)} {\sqrt{n\sum_{i=1}^{m}V(\gamma_i)}}\right),
   \label{eqir}
\end{equation}
where $\Gamma^{(m)}=(\gamma_1,\cdots,\gamma_m)$ is the vector of SINR of $m$ copies of the packet, $\gamma_{\mathrm{cc}}=\sum_{i=1}^{m} \gamma_i$ is the overall SINR after performing MRC at the BS,  $Q(x)=\frac{1}{\sqrt{2 \pi}} \int_{x}^{\infty}  e^{-\frac{u^2}{2}} du$ is the standard Q-function, $V(\gamma)=\left(1-(1+\gamma)^{-2} \right)\log^2_2(e)$ is the channel distortion \cite{polyanskiy2010channel}, $n$ is the packet length and $k$ is the information block length. In these formulations we refer to the ratio $R=k/n$ as the channel code rate\footnote{In practice we need to employ channel codes, such as BCH, Polar, or Reed-Muller codes, which are shown to perform close to the normal approximation benchmark \cite{shirvanimoghaddam2018short}.}. For the simplicity of analysis, we assume that all the packets, including the retransmission ones, have the same length, $n$. 
\vspace{-2ex}
\section{Reliability and Throughput Analysis}
\begin{figure}[t]
\centering
\includegraphics[width=0.85\columnwidth]{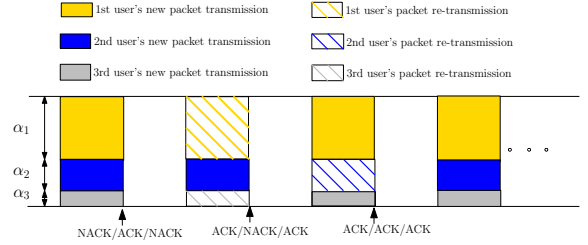}
 \vspace{-2ex}
\caption{NOMA-HARQ under various power splitting ratios $\alpha_{i}$.}
\label{fig:mnoma}
\end{figure}
Fig. \ref{fig:mnoma} shows the NOMA-HARQ with one retransmission when 3 users are sharing the same radio resource. As the BS sends individual feedback to each user after each time slot, each user's packet status will change differently from those of the other users. In this section, we propose a Markov model to evaluate the dynamics of the system and characterize the reliability and throughput.
 \vspace{-1ex}
\subsection{Markov Model}
\begin{figure}[t]
\centering
\includegraphics[width=0.95\columnwidth]{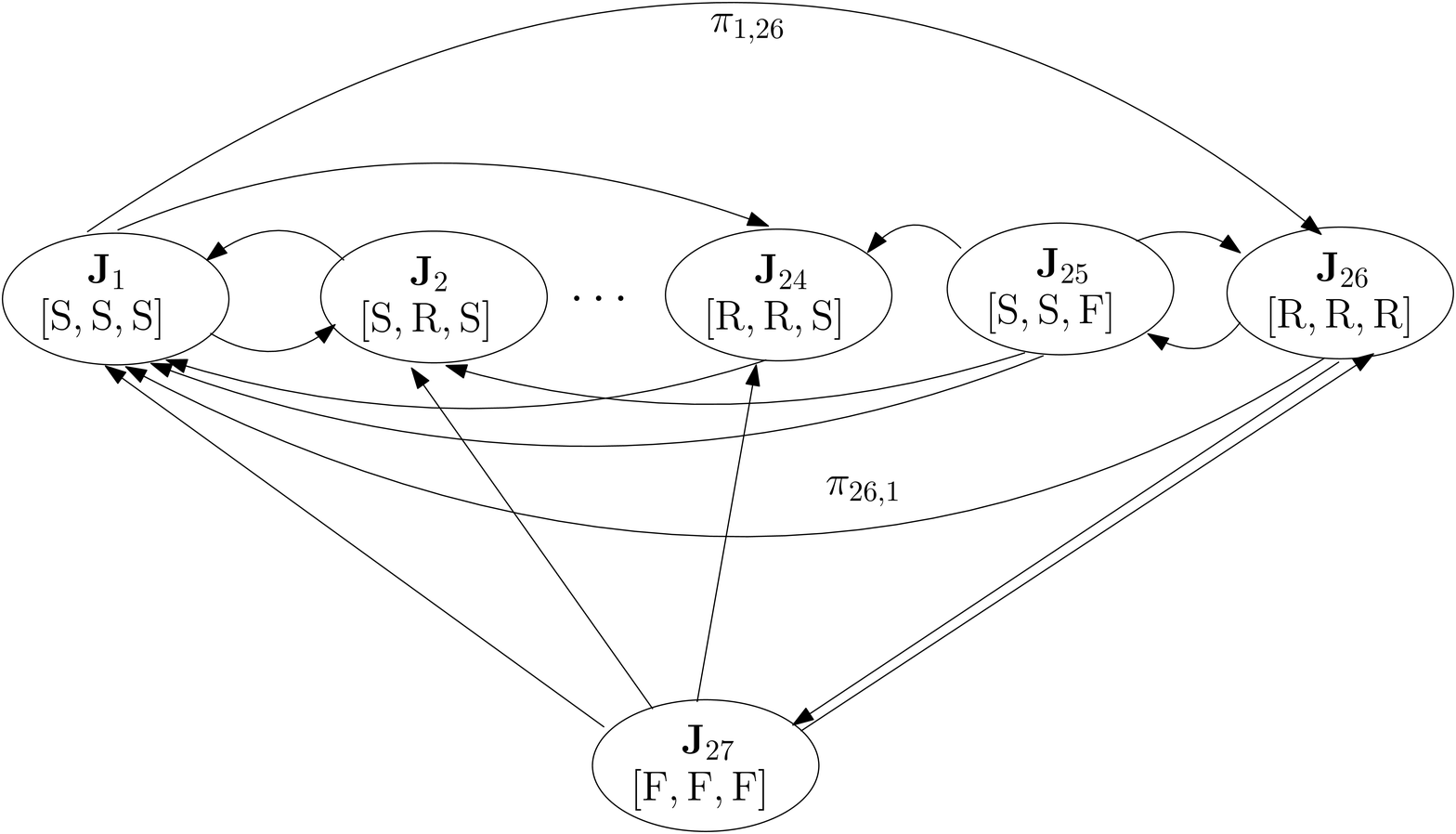}
 \vspace{-2ex}
\caption{Markov model for the 3-user NOMA-HARQ with one re-transmission. Transitions probabilities are omitted for the clarity of presentation.}
\label{fig:Markov}
\end{figure}

We use a Markov model, whose states are represented by a vector $\mathbf{J} = [J_1, J_2, \cdots,J_N]$, where $N$ is the number of active users that are sharing the same radio resource via NOMA, $L$ is the number of transmissions (i.e., the number of retransmissions is at most $L-1$), and $J_i \in \{\mathrm{S},\mathrm{R}_1, \mathrm{R}_2,\cdots,\mathrm{R}_{L-1},\mathrm{F}\}$ is the current state for the $i^{th}$ user. $\mathrm{F}$ refers to the failure state, where the packet failed the decoding after $L$ transmissions, $\mathrm{S}$ means that the packet succeeds and $\mathrm{R}_i$ means that the packet is being re-transmitted $i$ times. The total number of states in the Markov model is given by $(L+1)^{N}$. Fig. \ref{fig:Markov} shows the Markov model for a 3-user NOMA-HARQ with one allowed re-transmission, i.e., $N=3$ and $L=2$. For the ease of notation, we removed the subscript for the retransmission stage as $L=2$. It is important to note that when the system is currently at state $\mathrm{S}$ or $\mathrm{F}$, the next time slot will be used to transmit a new packet and depending on the feedback it receives from the BS, it may go to $\mathrm{S}$ or $\mathrm{R}$. The system will never go from state $\mathrm{S}$ to $\mathrm{F}$ directly, as it always performs the retransmission, whenever it receives a NACK.
\vspace{-2ex}
\subsection{Successive Interference Cancellation at the BS}
In the rest of the paper, we consider that $L=2$. That is, each user is allowed to retransmit the failed packet only once. This is mainly to keep the latency at minimum, which is also aligned with the recent proposals for communication among machines and robots in Industry 4.0 use cases \cite{popovski2019wireless}. 

Let $\mathcal{N}$ denote the set of all active users who are sharing the same radio resource via NOMA. Before starting SIC at the BS at the end of each time slot, we form two sets, where $\mathcal{R}(\mathbf{J})=\{j \in \mathcal{N}|J_{j}=\mathrm{R}\}$ denotes the set of users who are currently at state $\mathrm{R}$ and $\mathcal{F}(\mathbf{J})=\{j \in \mathcal{N}|J_{j}=\mathrm{F}\}$ denotes the set of users who are currently at state $\mathrm{F}$. It is then easy to show that the overall SINR for the $i^{th}$ user after transiting from its current state will be given by:
\begin{align}
     \gamma_{i}=\begin{cases}
        \displaystyle\frac{P_i}{{\displaystyle\sum_{{j} \in {\mathcal{N}}\setminus i}{{{P_{j}}}}}+1},& J_i \in \{\mathrm{F,S}\},\\
      \displaystyle {\frac{P_{i}}{{\displaystyle\sum_{{j} \in\{ {{{\mathcal{R}(\mathbf{J})}}\bigcup{\mathcal{F}(\mathbf{J})}}\}\setminus i}{{{P_{{j}}}}}}+1}+ \displaystyle\frac{P_{i}}{{\displaystyle\sum_{{j} \in {\mathcal{N}}\setminus i}{{{P_{{j}}}}}}+1}},& J_i=\mathrm{R},\\
     \end{cases}
     \label{eeee}
\end{align}
where for the user at state $\mathrm{R}$, the BS will have two copies of its packet, therefore it performs MRC. The BS then finds the user with the highest SINR as follows:
\begin{align}
    I_1=\arg\max_i \gamma_i.
\end{align}
The BS then attempts decoding the user with index $I_1$ in the first stage of SIC. Let $I_{\ell}$, for $1<\ell\le N$, denote the user index with the highest SINR after completing the $(\ell-1)^{th}$ stage of SIC; that is all the previous $(\ell-1)$th stages of SIC led to a successful decoding. Let $\mathcal{N}_{\ell}(\mathbf{J}):=\{I_1,\cdots,I_{\ell}\}$ denotes the set of $\ell$ users which have been decoded in the first $\ell^{th}$ stages of SIC. After completing the $(\ell-1)^{th}$ stage of SIC, the SINR of the remaining users, i.e., for $j\in\mathcal{N}^c_{\ell-1}$, where $\mathcal{N}^c_{\ell}(\mathbf{J}) = \mathcal{N}-\mathcal{N}_{\ell}(\mathbf{J})$, can be calculated as follows:
\begin{align}
     \gamma^{(\ell)}_{j}=\begin{cases}
        \displaystyle\frac{P_j}{{\displaystyle\sum_{{w} \in {\mathcal{N}_{\ell-1}^c}(\mathbf{J})\setminus j}{{{P_{w}}}}}+1},& \hspace{-3ex}J_j \in \{\mathrm{F,S}\},\\
        \displaystyle\frac{P_j}{{\displaystyle\sum_{{w} \in {\mathcal{N}_{\ell-1}^c(\mathbf{J})}\setminus j}{{{P_{w}}}}}+1}+\displaystyle\frac{P_j}{{\displaystyle\sum_{w \in \mathcal{R}^c_{\ell-1}(\mathbf{J})\setminus j}{{{P_w}}}}+1},& J_j=\mathrm{R},\\
     \end{cases}
     \label{eq:SINR-SIC-J}
\end{align}
where $\mathcal{R}^c_{\ell-1}(\mathbf{J}):={\mathcal{F}(\mathbf{J})\bigcup\{\mathcal{R}(\mathbf{J})-\mathcal{N}_{\ell-1}(\mathbf{J})\}}$. It is important to note that when $J_j=\mathrm{R}$, there would be two copies of the packet at the BS. For the retransmitted packet, users' packets which have not been decoded yet, i.e., ${\mathcal{N}_{\ell-1}^c(\mathbf{J})}\setminus j$, will interfere with it. For the original packet, all users which are currently at state $\mathrm{F}$ and those at state $\mathrm{R}$ which have not been decoded yet, will interfere with it. In other words, all users in ${\mathcal{F}(\mathbf{J})\bigcup\{\mathcal{R}(\mathbf{J})-\mathcal{N}_{\ell-1}(\mathbf{J})\}}\setminus j$, will interfere with the original packet. As the BS performs MRC, the overall SINR for the user $j$ currently at state $\mathrm{R}$ can be calculated as in (\ref{eq:SINR-SIC-J}). The BS then finds the index of the user to be decoded in the $\ell^{th}$ stage of SIC as follows:
\begin{align}
I_{\ell}=\arg\max_j \gamma^{(\ell)}_{j}.
\end{align}
The BS performs SIC and proceeds to the next stage, only if the decoding succeeded in the previous stage. Otherwise, it terminates and the BS will send NACK to the remaining users.  It is important to note that the BS can uniquely determine the SIC decoding order at each state $\textbf{J}$, that is $\mathcal{N}_N(\mathbf{J})=\{I_1, I_2, \cdots, I_N\}$, is known at the BS and calculated by (\ref{eq:SINR-SIC-J}).

\subsection{State Transition Probabilities}
The  following  lemma characterizes the probability of transiting from state $\mathbf{J}$ to state $\mathbf{J}'$, denoted by $\pi_{\mathbf{J}\rightarrow \mathbf{J}'}$.  
\begin{lemma}
For the NOMA-HARQ with maximum one re-transmission, the probability of transiting from state $\mathbf{J}$ to $\mathbf{J}'$, denoted by $\pi_{\mathbf{J}\rightarrow \mathbf{J}'}$, can be calculated as follows:
\begin{align}
     \pi_{\mathbf{J}\rightarrow \mathbf{J}'}=\begin{cases}
        0,& \exists i:J_i\in\{\mathrm{S,F}\}, J'_i=\mathrm{F},\\
       0,& \exists i,j: I_i<I_j, J'_{I_i}\in\{\mathrm{F,R}\}, J'_{I_j}=\mathrm{S},\\
       \prod_{w=1}^{m}q_{I_w},&\mathrm{otherwise,}
     \end{cases}
     \label{eq:transprob}
\end{align}
where $[I_1, \cdots, I_N]$ is the SIC decoding order at state $\mathbf{J}$, 
\begin{align}
    m=\min\left\{i\left|J'_{I_i}\right.\in\{\mathrm{R,F}\}\right\},
\end{align} and
\begin{equation}
     q_{{I_i}}=\begin{cases}
         1- \epsilon\left(\gamma^{(i)}_{I_i}\right),&  J'_{I_i} =\mathrm{S},\\
     \epsilon\left(\gamma^{(i)}_{I_i}\right), & J'_{I_i}=\{\mathrm{R,F}\},
     \end{cases}
     \label{eq:statetransuser}
\end{equation}
where $\epsilon\left(\gamma^{(i)}_{I_i}\right)$ can be calculated by (\ref{cc}) for CC-HARQ.
\end{lemma}
 \begin{IEEEproof}
See Appendix A for the proof.
\end{IEEEproof}

Let $\mathcal{J}=\{\mathbf{J}_1, \cdots, \mathbf{J}_{3^N}\}$ denote the set of all states for NOMA-HARQ with one retransmission and let $\Pi$ denotes the state transition matrix for NOMA-HARQ with one re-transmission, then it can be constructed as follows:
\begin{equation}
    \Pi=
\begin{bmatrix}
\pi_{1,1} &\pi_{1,2}& \dots &\pi_{1,3^{N}}& \\
\pi_{2,1} &\pi_{2,2} &\dots& \pi_{2,3^{N}} &\\
\vdots  & \vdots & \vdots &  \vdots \\
\pi_{3^{N}, 1} &\pi_{3^{N}, 2}& \dots& \pi_{3^{N}, 3^{N}}& \\ 
\end{bmatrix},
\end{equation}
where $\pi_{i,j}$ is probability of transiting from state $i$ to $j$ and can be calculated according to Lemma 1. The state stationary distribution, denoted by $P_{\mathrm{stat}}$, can be calculated explicitly through obtaining the eigen vector associated with the unity eigenvalue of matrix $\Pi^{\mathrm{T}}$ as follows:
\begin{equation}
    \Pi^{\mathrm{T}} P_{\mathrm{stat}} = P_{\mathrm{stat}},
\end{equation}
where superscript $^\mathrm{T}$ is the matrix transpose operation. 
\vspace{-2.5ex}
\subsection{Packet Error Rate}
The following lemma characterizes the packet error rate for the NOMA-HARQ with one retransmission. 
\begin{lemma}
\label{lemmaper}
Let $\mathcal{J}=\{\mathbf{J}_1, \cdots, \mathbf{J}_{3^N}\}$ denote the set of all states for NOMA-HARQ with one retransmission and $P_{\mathrm{stat}} = [p_1,\cdots, p_{3^N}]^{\mathrm{T}}$ denote  the state stationary distribution of the respective Markov model,  where $p_w$ denote the state stationary probability of $\mathbf{J}_w$. Then, the packet error rate for $i^{th}$ user, denoted by $e_i$, is given by:
\begin{align}
    e_i=\sum_{w\in {{\mathcal{F}}_{i}}}{{{p}_{w}}}+\sum_{w\in {{{{\mathcal{R}}}}_{i}}}{{{p}_{w}}\left( \sum_{j\in {{\mathcal{F}}_{i}}}{{{\pi }_{w,j}}} \right)},
    \label{eqper1}
\end{align}
where ${{\mathcal{F}}_{i}}=\{\mathbf{J} \in \mathcal{J}|J_{i} =\mathrm{F}\}$ and ${{{\mathcal{R}}}_{i}}=\{\mathbf{J} \in \mathcal{J}|J_{i} =\mathrm{R}\}$.
\end{lemma}

\begin{IEEEproof}
See Appendix B.
\end{IEEEproof}
\vspace{-2.5ex}
\subsection{Throughput Analysis}
The following lemma characterizes the delay distribution for each user in the NOMA-HARQ with one retransmission.
\begin{lemma}
\label{lemmadelay}
Let $D^{(i)}$ denotes the number of packets (including re-transmissions) that user $i$ needs to transmit to deliver $M$ information packets. The probability mass function ($\mathrm{pmf}$) of $D^{(i)}$ is given by:
\begin{align}
    \mathrm{Prob}\left\{D^{(i)}=2M-j\right\}=\dbinom{M}{j}{p^{(i)}_\mathrm{s}}^j\left(1-p^{(i)}_\mathrm{s}\right)^{M-j},
    \label{eqbinomial}
\end{align}
where $j\in\{0,\cdots,M\}$ and
\begin{align}
p^{(i)}_\mathrm{s}=\sum_{w\in {\mathcal{S}_i \bigcup\mathcal{F}_i}}{{{p_w}}\sum_{{j} \in {\mathcal{S}_i}}{{{\pi }_{w,j}}}},
\label{eqpsuccess}
\end{align}
and $\mathcal{S}_i=\{\mathbf{J} \in \mathcal{J}|{J}_{i}=\mathrm{S}\}$.
\end{lemma}

\begin{IEEEproof}
See Appendix C.
\end{IEEEproof}
\begin{figure*}[t]
\centering
\includegraphics[width=0.66\columnwidth]{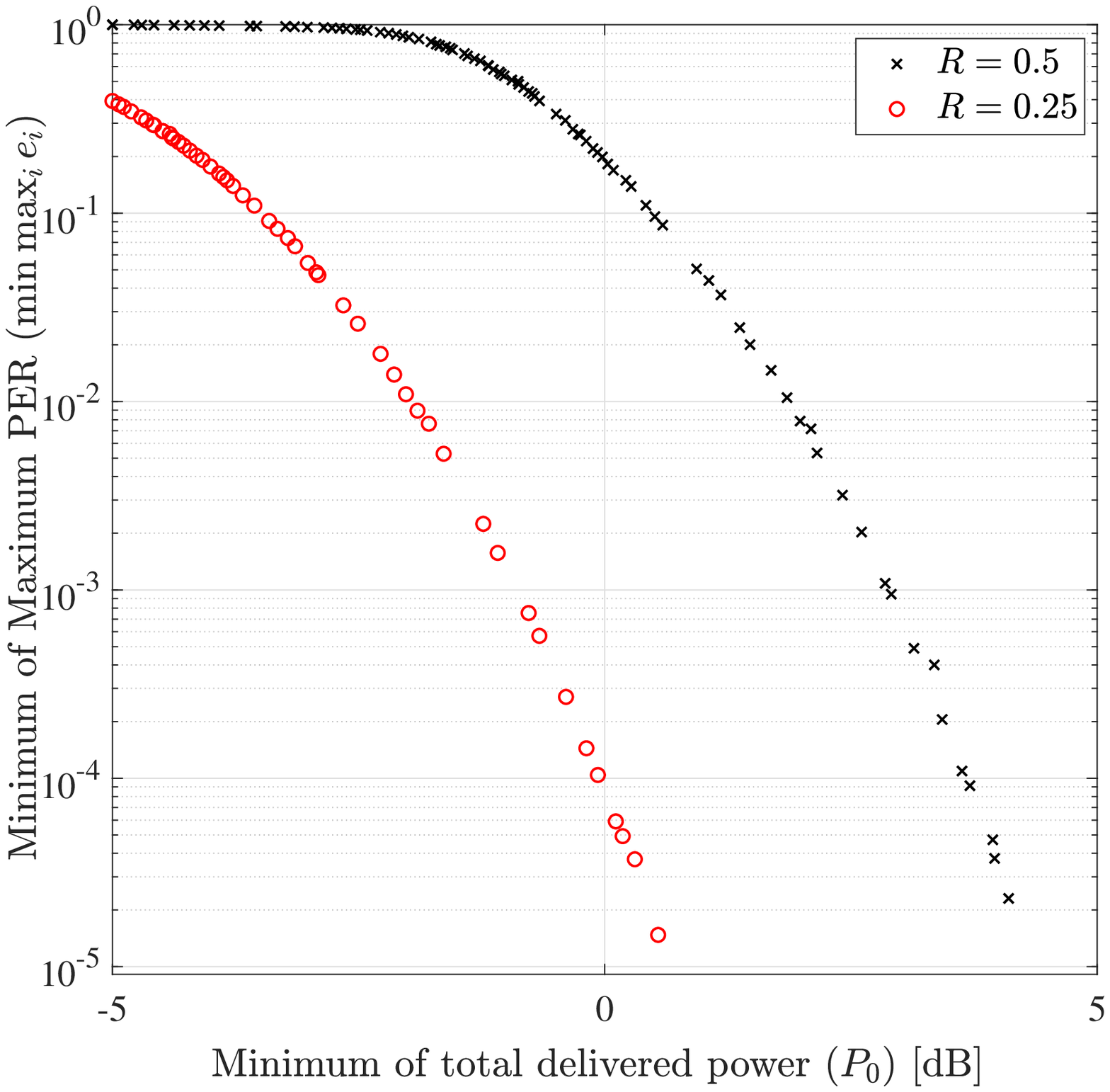}
\includegraphics[width=0.66\columnwidth]{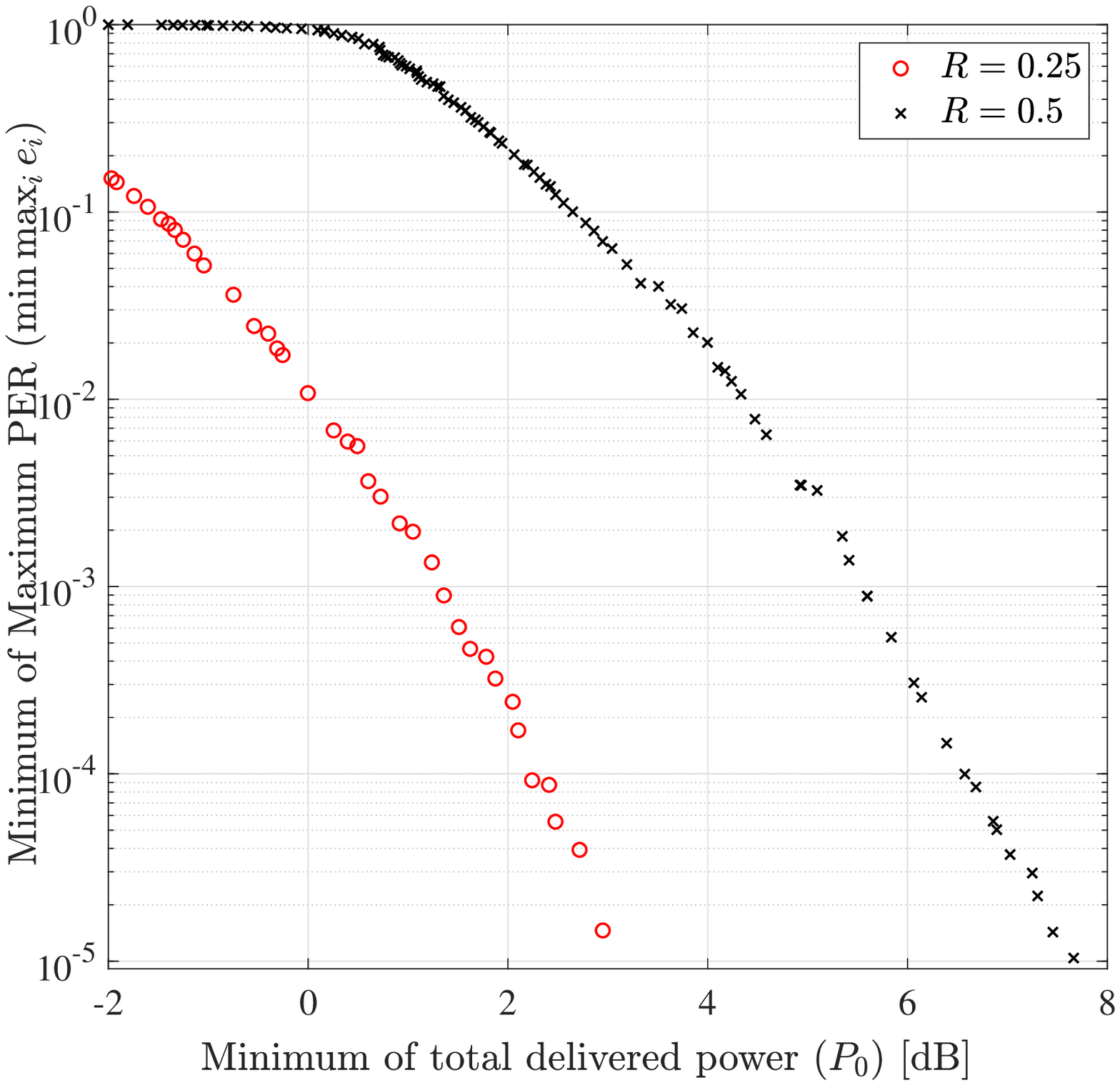}
\includegraphics[width=0.66\columnwidth]{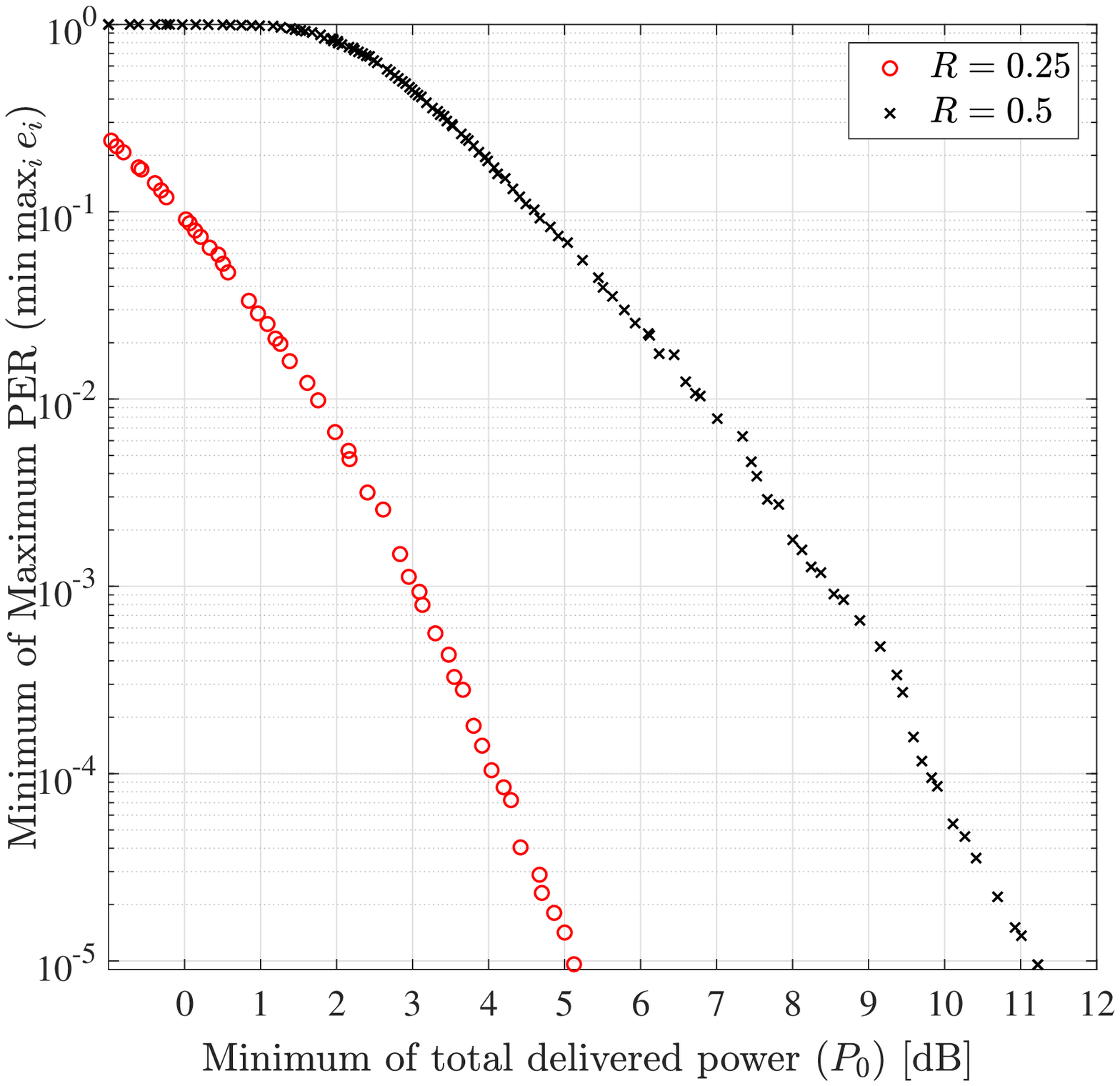}\\
\footnotesize a) $N=3$ ~~~~~~~~~~~~~~~~~~~~~~~~~~~~~~~~~~~~~~~~~~~~~~b) $N=4$ ~~~~~~~~~~~~~~~~~~~~~~~~~~~~~~~~~~~~~~~~~~~~~~~c) $N=5$\\
\caption{Pareto front for bi-objective optimization (\ref{Opt1}) for NOMA-HARQ with one re-transmission when $n=100$ and a) $N=3$, b) $N=4$, and c) $N=5$ at different code rates $R=0.5$ and $R=0.25$.}
\vspace{-3ex}
\label{fig:ParetoFront}
\end{figure*}
\begin{table}[t]
    \centering
        \caption{Optimum power splitting ratios for the energy constrained scenario, i.e., optimization problem (\ref{Opt1}) when $n=100$.}
        \scriptsize
\begin{tabular}{|c|c|c|c|c|c|c|c|}
\hline
\multirow{2}{*}{$\min\max_i{e_i}$} & \multicolumn{5}{c|}{Power Splitting ratios} & \multirow{2}{*}{$P_0$ [dB]} & \multirow{2}{*}{$R=\frac{k}{n}$} \\
\cline{2-6}
& $\alpha_1$ & $\alpha_2$ & $\alpha_3$ & $\alpha_4$ & $\alpha_5$& &\\
\hline
$7.5\times10^{-3}$ & 0.29 & 0.35 & 0.36 & - & - & -2.02&\multirow{4}{*}{0.25} \\
\cline{1-7}
$10^{-3}$ & 0.29 & 0.35& 0.36 & - & - & -0.77 &\\
\cline{1-7}
$10^{-4}$& 0.28 & 0.34 & 0.38 & - & - & -0.07 &\\
\cline{1-7}
$8.85\times10^{-6}$& 0.27 & 0.34 & 0.39 & - & - & 0.69 &\\
\hline
$10^{-2}$ & 0.27 & 0.32 & 0.41 & - & - & 1.85 & \multirow{4}{*}{0.5}  \\
\cline{1-7}
$10^{-3}$ & 0.25 & 0.33 & 0.42 & - & - & 2.85 & \\
\cline{1-7}
$10^{-4}$ & 0.24 & 0.33 & 0.43 & - & - & 3.63 & \\
\cline{1-7}
$8.97\times10^{-6}$ & 0.23 & 0.32 & 0.45 & - & - & 4.38 &\\
\hline
$10^{-2}$ & 0.20 & 0.24 & 0.25 & 0.31 & - & 0 & \multirow{4}{*}{0.25}  \\
\cline{1-7}
$9\times10^{-4}$ & 0.20 & 0.23 & 0.25 & 0.32 & - & 1.36 &\\
\cline{1-7}
$9.24\times10^{-5}$ & 0.19 & 0.22 & 0.26 & 0.33 & - & 2.24 &\\
\cline{1-7}
$8.97\times10^{-6}$ & 0.18 & 0.22 & 0.25 & 0.35 & - & 3.18 &\\
\hline
$10^{-2}$ & 0.17 & 0.21 & 0.27 & 0.34 & - & 4.33 & \multirow{4}{*}{0.5} \\
\cline{1-7}
$8.8\times10^{-4}$ & 0.15 & 0.22 & 0.28 & 0.35 & - & 5.59 & \\
\cline{1-7}
$9.974\times10^{-5}$ & 0.14 & 0.21 & 0.28 & 0.37 & - & 6.57 & \\
\cline{1-7}
$10^{-5}$ & 0.13 & 0.21 & 0.28 & 0.38 & - & 7.66 &\\
\hline
$9.8\times10^{-3}$ & 0.15 & 0.17 & 0.19 & 0.23 & 0.26 & 1.76 & \multirow{4}{*}{0.25}\\
\cline{1-7}
$9.3\times10^{-4}$ & 0.14 & 0.16 & 0.19 & 0.24 & 0.27 & 3.09 &\\
\cline{1-7}
$10^{-4}$ & 0.13 & 0.16 & 0.2 & 0.24 & 0.27 & 4.04 &\\
\cline{1-7}
$9.596\times10^{-6}$ & 0.13 & 0.15 & 0.19 & 0.24 & 0.29 & 5.12 & \\
\hline
$10^{-2}$ & 0.11 & 0.15 & 0.2 & 0.24 & 0.3 & 6.78 & \multirow{4}{*}{0.5}\\
\cline{1-7}
$9.1\times10^{-4}$ & 0.1 & 0.14 & 0.21 & 0.24 & 0.31 & 8.54 &\\
\cline{1-7}
$9.5\times10^{-5}$ & 0.07 & 0.09 & 0.15 & 0.18 & 0.51 & 9.83 &\\
\cline{1-7}
$9.55\times10^{-6}$ & 0.07 & 0.1 & 0.14 & 0.18 & 0.51 & 11.22 &\\
\hline
\end{tabular}
\label{tab:obj1}
\end{table} 
\vspace{-1ex}
Using Lemma \ref{lemmaper} and Lemma \ref{lemmadelay}, the throughput for user $i$, denoted by $\eta_i$, can be calculated as follows:
\begin{equation}
  \eta_i= R\frac{1-e_i}{p^{(i)}_\mathrm{s}+2(1-p^{(i)}_\mathrm{s})},  
  \label{eqthroughput}
\end{equation}
where $R=k/n$ is the code rate. 
\vspace{-2ex}
\section{Optimization of Power Splitting Ratios for NOMA-HARQ}
In this section, we define two different optimization problems. We first consider a power constrained scenario, in which we jointly minimize the maximum packet error rate among users and the total received power at the BS. This scenario is mainly suitable for mMTC. Second, we consider the throughput maximization problem for the reliable transmission of packets, in which each user's packet error rate must satisfy a reliability constraint. This will be mostly suitable for the URLLC scenario. 
\vspace{-2ex}
\subsection{The Power Constrained Scenario}
We consider a bi-objective optimization problem, where we jointly minimize the maximum packet error rate among all users and the required total received power at the BS, $P_0$. In particular, for the information block length $k$, codeword length $n$, and the number of active users $N$, the optimization problem is summarized as follows:
\begin{align}
    &\min_{\{\alpha_1,\cdots, \alpha_N\}} \left\{\max_{i\in[1,N]}\{e_i\},P_0\right\} ,\label{Opt1}\\
    \mathrm{s.t.} ~~~ &\sum_{i=1}^N{{{\alpha}_{i}}}=1.\notag
\end{align}
The constraint specifies that the total delivered power at the receiver is $P_0$. It is important to note that for such a nontrivial multi-objective optimization problem, no single solution exists that simultaneously optimizes both objectives. This optimization problem can be numerically solved for different value of $k$, $n$, $P_0$, and $N$. In this paper, we apply the genetic algorithm inspired by evolutionary biology and introduced by Holland \cite{holland1975adaptation} in solving optimization problems. This algorithm utilizes selection, crossover, and mutation operators to come as close as possible to the optimal solution \cite{eiben2003introduction}. Table \ref{tab:obj1} shows the optimal power splitting ratios and the respective optimal packet error rate for different values of $N$, code rate $R$, and received power $P_0$ when $n=100$. 

The Pareto Front obtained for optimization problem (\ref{Opt1}) for different values of $N$ and code rates is shown in Fig. \ref{fig:ParetoFront}. As it can be seen, the system achieves a higher power efficiency at the given target PER when a lower code rate is chosen.  It is also clear from Fig. \ref{fig:ParetoFront} that when the number of active users increases, the system can achieve the same level of PER with the same code rate $R$ at the expense of increased delivered power at the receiver. 
\begin{figure*}[t]
\centering
\includegraphics[width=0.66\columnwidth]{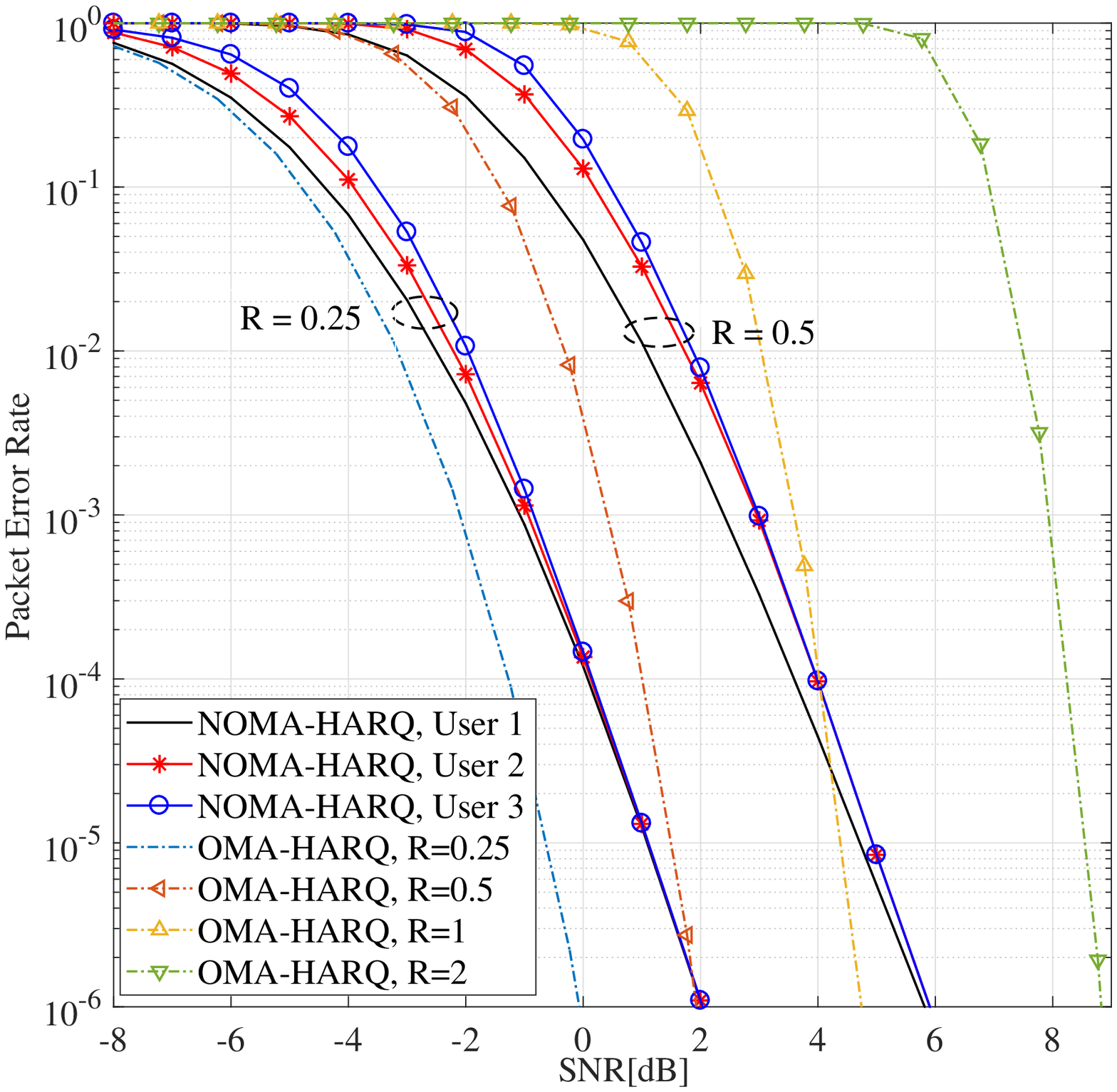}
\includegraphics[width=0.66\columnwidth]{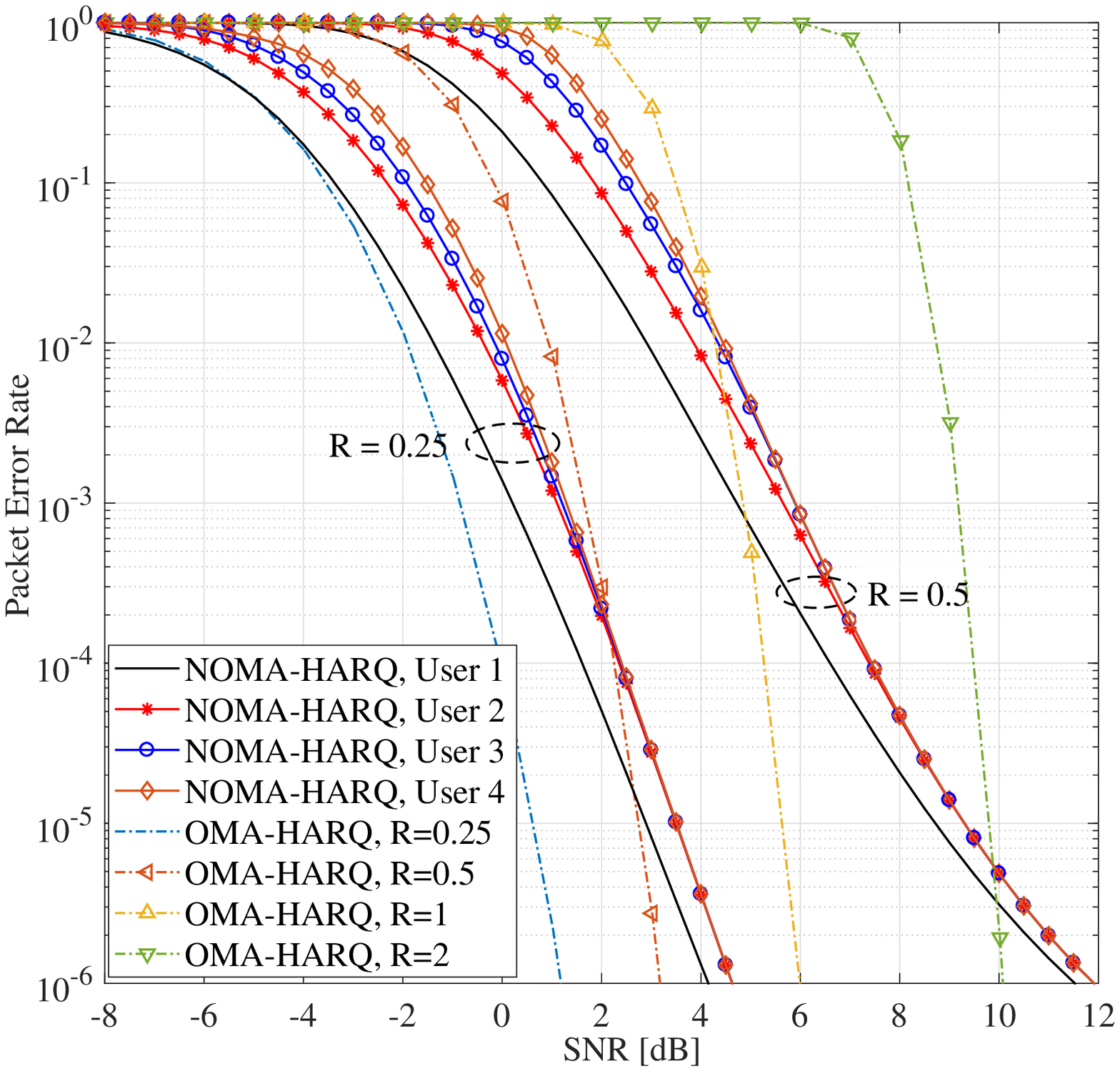}
\includegraphics[width=0.66\columnwidth]{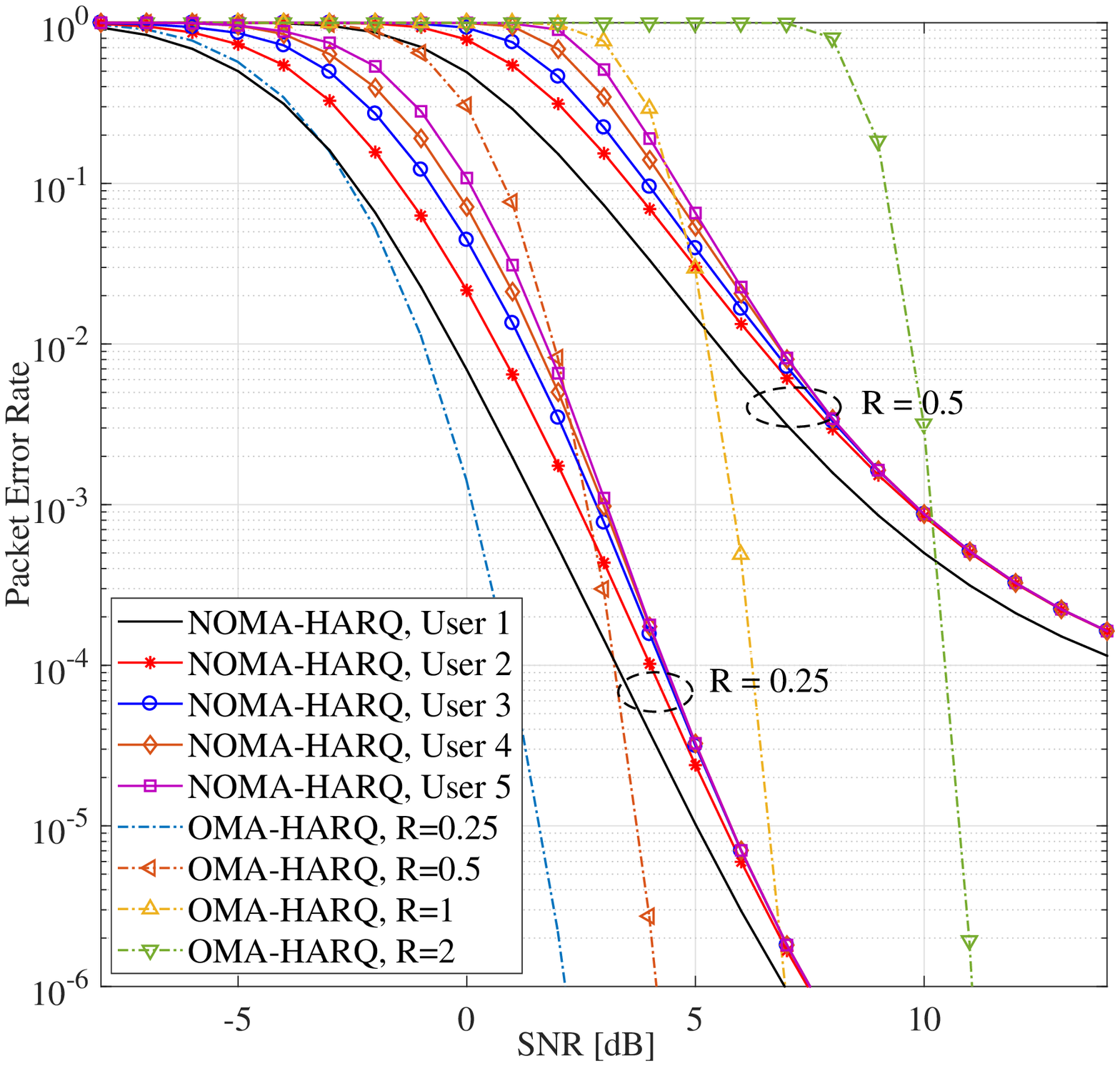}
\\
\footnotesize a) $N=3$ ~~~~~~~~~~~~~~~~~~~~~~~~~~~~~~~~~~~~~~~~~~~~~~b) $N=4$ ~~~~~~~~~~~~~~~~~~~~~~~~~~~~~~~~~~~~~~~~~~~~~~~c) $N=5$
 \vspace{-2ex}
\caption{Packet error rate versus SNR for NOMA-HARQ and OMA-HARQ with one retransmission, when $n=100$ and two different values of channel code rate, $R=0.25$ and $R=0.5$, where the number of active users is a) $N=3$, b) $N=4$, and c) $N=5$. The power splitting ratios of NOMA-HARQ were obtained from Table \ref{tab:obj1} for the target error probability of $10^{-2}$.}
\label{fig:PER-COOR}
\end{figure*}

\begin{figure*}[t]
\centering
\includegraphics[width=0.66\columnwidth]{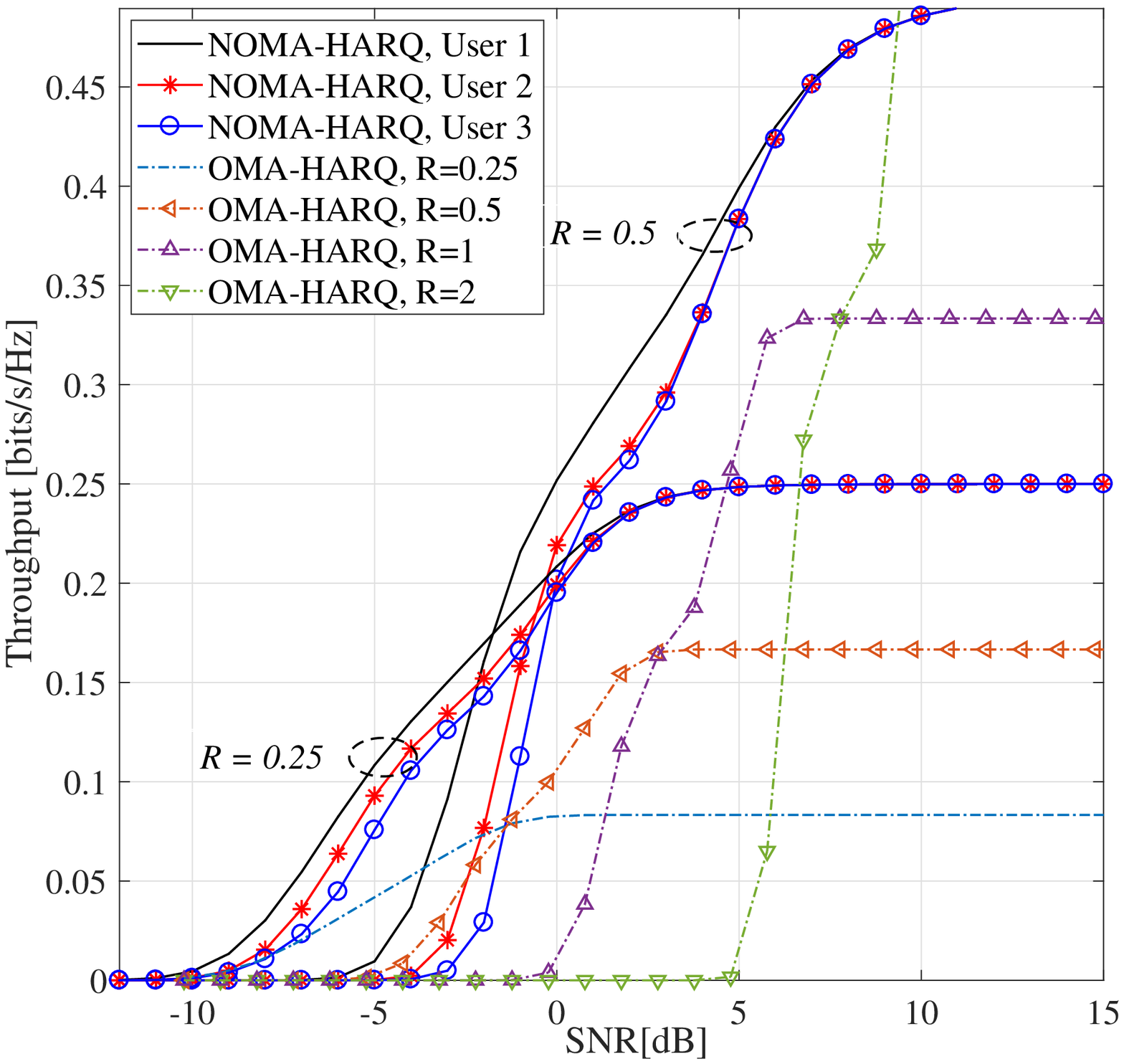}
\includegraphics[width=0.66\columnwidth]{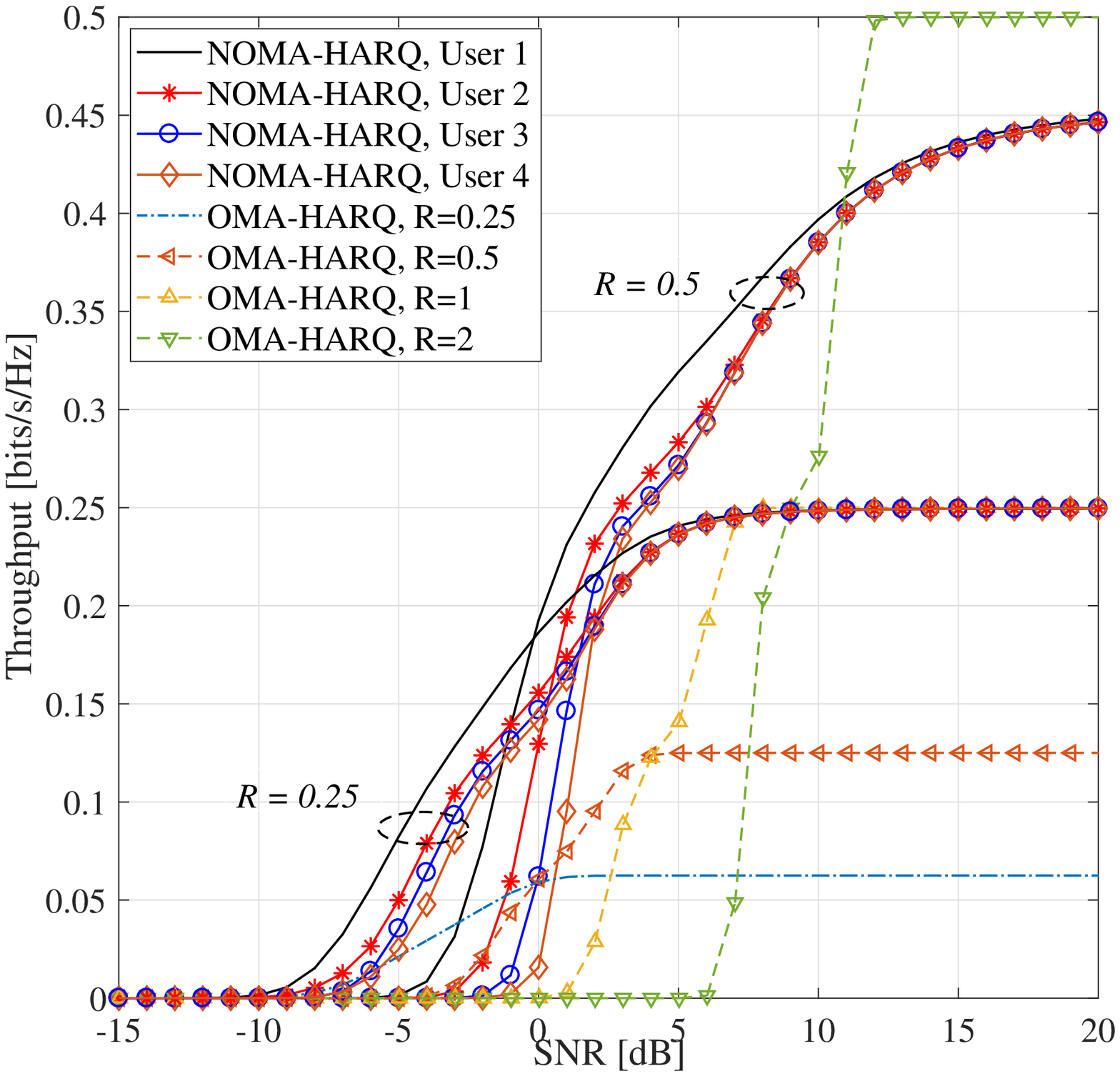}
\includegraphics[width=0.66\columnwidth]{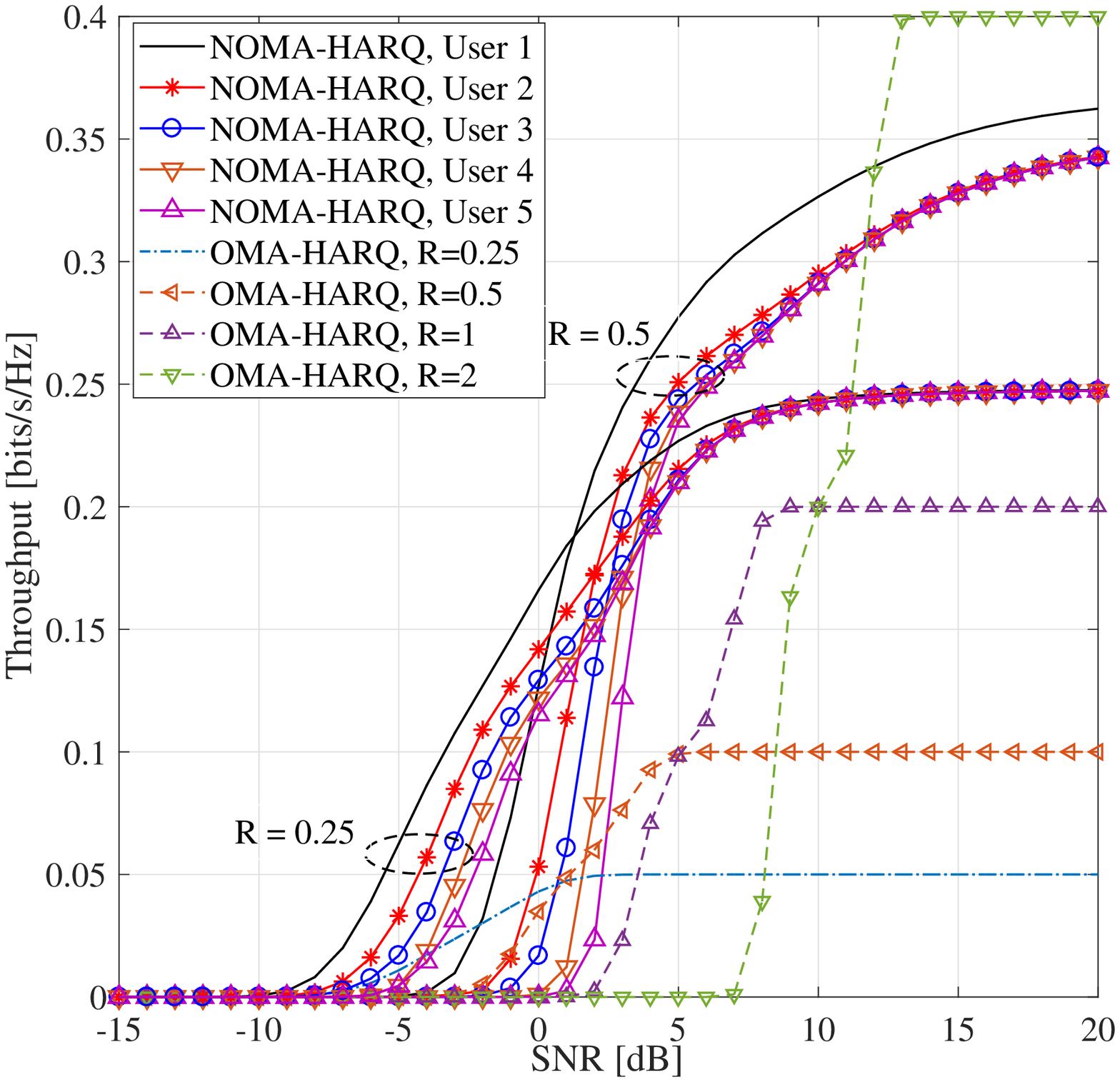}
\\
\footnotesize a) $N=3$ ~~~~~~~~~~~~~~~~~~~~~~~~~~~~~~~~~~~~~~~~~~~~~~b) $N=4$ ~~~~~~~~~~~~~~~~~~~~~~~~~~~~~~~~~~~~~~~~~~~~~~~c) $N=5$
\caption{Throughput versus SNR for NOMA-HARQ and OMA-HARQ with one retransmission for different values of channel code rate, $R=0.25$ and $R=0.5$, when $n=100$ and the target packet error rate for all users is $e_t=10^{-2}$. The power splitting ratios of NOMA-HARQ were obtained from Table \ref{tab:obj1} for the target error probability of $10^{-2}$.}
\label{fig:Throughput-Coor}
 \vspace{-2em}
\end{figure*}

Fig. \ref{fig:PER-COOR} shows the packet error rate of both NOMA-HARQ and OMA-HARQ schemes versus the received SNR at the BS for all users. The optimal power splitting ratios for NOMA-HARQ scheme were obtained from Table \ref{tab:obj1} for the target error probability of $10^{-2}$. As can be seen in Fig. \ref{fig:PER-COOR}-a when $N=3$ all the users with NOMA-HARQ scheme achieve the desired level of reliability with only a slight difference in the required SNR. This property persists when a higher code rate is chosen. This is consistent with the results shown in Fig. \ref{fig:PER-COOR}-b for $N=4$ and \ref{fig:PER-COOR}-c for $N=5$. It is important to note that when the number of users increases and when a higher code rate is chosen, the PER curve shows an error floor which is mainly due to the error propagation effect in SIC. For the sake of comparison, we also consider OMA-HARQ and show its PER performance in Fig. \ref{fig:PER-COOR}. In OMA-HARQ, each time slot will be occupied by one user only. We further assume that the users allow to perform one re-transmission provided that their original transmission failed. We also assume that the average total received power at the BS per information packet is the same as that for NOMA-HARQ in order to have a fair comparison. As can be seen in Fig. \ref{fig:PER-COOR}, OMA-HARQ outperforms NOMA-HARQ in terms of PER which is obvious since in OMA-HARQ only one user uses the channel and there is no interference among users. However, in NOMA-HARQ scheme, users share the same channel and the user interference will degrade the PER performance.

We show the throughput performance of both NOMA-HARQ and OMA-HARQ schemes in Fig. \ref{fig:Throughput-Coor} for different code rates and number of users when the target error probability is set to $10^{-2}$. As can be seen in this figure, for medium SNRs the user with a higher power ratio achieves a higher throughput in NOMA-HARQ scheme while for high SNRs the throughput of all users converge to the maximum throughput. It is important to note that by using NOMA-HARQ with only one retransmission, all users can achieve the desired level of reliability or throughput when the power splitting ratios are well chosen. When the number of users is high, for example $N=5$, the desired level of PER may not be achieved when the code rate is high (see Fig. \ref{fig:PER-COOR}-c); therefore one needs to decrease the code rate at the expense of reduced throughput. It should be also noted that under the same rate and average received power at the BS, NOMA-HARQ outperforms OMA-HARQ in terms of throughput. This is mainly because in OMA-HARQ users are transmitting separately and need to wait for the other users to complete their transmissions, which might also include the retransmission. As reflected in this figure, even if the rate $R$ increases, which increase the PER, OMA-HARQ cannot generally achieve the same throughput as NOMA-HARQ specially in low-to-moderate SNRs. For high SNRs, one can also choose a higher rate for NOMA-HARQ, so it can achieve a higher throughput while satisfying the desired level of reliability. 
\vspace{-2ex}
\subsection{The Reliability Constrained Scenario}
The problem of finding the minimum required block length $n$, which guarantees the target maximum packet error rate among all users at the given SNR can be formulated as follows:
\begin{align}
    & \min_{\{\alpha_1,\cdots,\alpha_N\}} n,\\
    \mathrm{s.t.} ~~~& \sum_{i=1}^N{{{\alpha}_{i}}}=1, \notag\\
    & \max_{i\in[1,N]} \{e_i\}\le e_t,\notag
    \label{eq:Opt2}
\end{align}
where $e_t$ is the target packet error rate. 

This optimization problem is solved using an iterative algorithm. In particular, for a given target maximum packet error rate $e_t$ among all users at a given SNR, the algorithm is started with $n=k+1$ and uses the Genetic algorithm to find the optimum values of $\{\alpha_1,…,\alpha_N\}$ which minimizes $ \max_{i\in[1,N]} \{e_i\}$. If the maximum PER is larger than the target PER $e_t$, $n$ is increased by $1$ and we run the Genetic algorithm again to find the optimum values of $\{\alpha_1,…,\alpha_N\}$. The block length $n$ will be increased if the PER constraint cannot be met. The algorithm will stop once the PER constrain is met and the block length will be reported as the minimum required block length.

Table \ref{tab:Opt2} shows the optimal power splitting ratios for different target PERs and their respective minimum required block length. As can be seen in this table, when the block length $n$ is larger, a larger number of users can simultaneously achieve the desired level of reliability.

\begin{table}[t]
    \centering
        \caption{Optimum power splitting ratios for the reliability constraint system (\ref{eq:Opt2}), when SNR$=0$ dB and $k=50$.}
                \scriptsize
\begin{tabular}{|c|c|c|c|c|c|c|}
\hline
\multirow{2}{*}{$e_t$} & \multicolumn{5}{c|}{Power Splitting ratios} & \multirow{2}{*}{Blocklength(n)}  \\
\cline{2-6}
& $\alpha_1$ & $\alpha_2$ & $\alpha_3$ & $\alpha_4$ & $\alpha_5$& \\
\hline
$9.7\times10^{-3}$ & 0.28 & 0.33 & 0.39 & - & - & 130  \\[0.7ex]
\hline
$10^{-3}$ & 0.28 & 0.33& 0.39 & - & - & 149   \\[0.7ex]
\hline
$10^{-4}$& 0.28 & 0.34 & 0.38 & - & - & 166   \\[0.7ex]
\hline
$10^{-5}$& 0.28 & 0.34 & 0.38 & - & - & 181  \\[0.7ex]
\hline
$10^{-2}$ & 0.20 & 0.23 & 0.27 & 0.3 & - & 176  \\[0.7ex]
\hline
$10^{-3}$ & 0.20 & 0.24 & 0.27 & 0.29 & - & 203  \\[0.7ex]
\hline
$10^{-4}$ & 0.21 & 0.24 & 0.26 & 0.29 & - & 227   \\[0.7ex]
\hline
$10^{-5}$ & 0.20 & 0.24 & 0.27 & 0.29 & - & 248 \\[0.7ex]
\hline
$10^{-2}$ & 0.16 & 0.18 & 0.20 & 0.22 & 0.24 & 223   \\[0.7ex]
\hline
$9.7\times10^{-4}$ & 0.16 & 0.18 & 0.20 & 0.22 & 0.24 & 259  \\[0.7ex]
\hline
$10^{-4}$ & 0.16 & 0.18 & 0.20 & 0.22 & 0.23 & 289   \\[0.7ex]
\hline
$10^{-5}$ & 0.16 & 0.18 & 0.20 & 0.22 & 0.24 & 316 \\[0.7ex]
\hline
\end{tabular}
\label{tab:Opt2}
\end{table}
\begin{figure}[t]
\centering
\includegraphics[width=0.8\columnwidth,height=6cm]{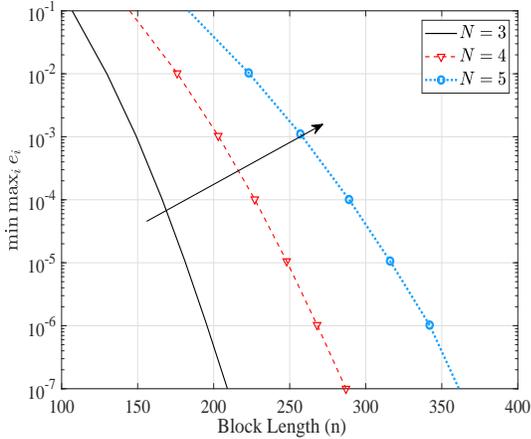}
 \vspace{-3ex}
\caption{The Minimum of maximum PER among all users versus block length $n$ for different numbers of users when the received SNR at the BS is $0$dB and $k=50$.}
\label{fig:PER-Opt2}
\end{figure}
Fig. \ref{fig:PER-Opt2} shows the packet error rate versus the block length $n$, for different number of active users, when the total received SNR at the BS is $0$ dB and the information block length $k$ is 50. As can be seen in this figure, one can easily control the maximum packet error rate among all the users by choosing the proper power splitting ratios and the block length. In particular, for a given information block length $k$, NOMA-HARQ with only one retransmission can simultaneously satisfy the reliability requirement of a larger number of users, when a larger  block length and accordingly a lower code rate are chosen.

\begin{figure*}[t]
\centering
\includegraphics[width=0.8\textwidth]{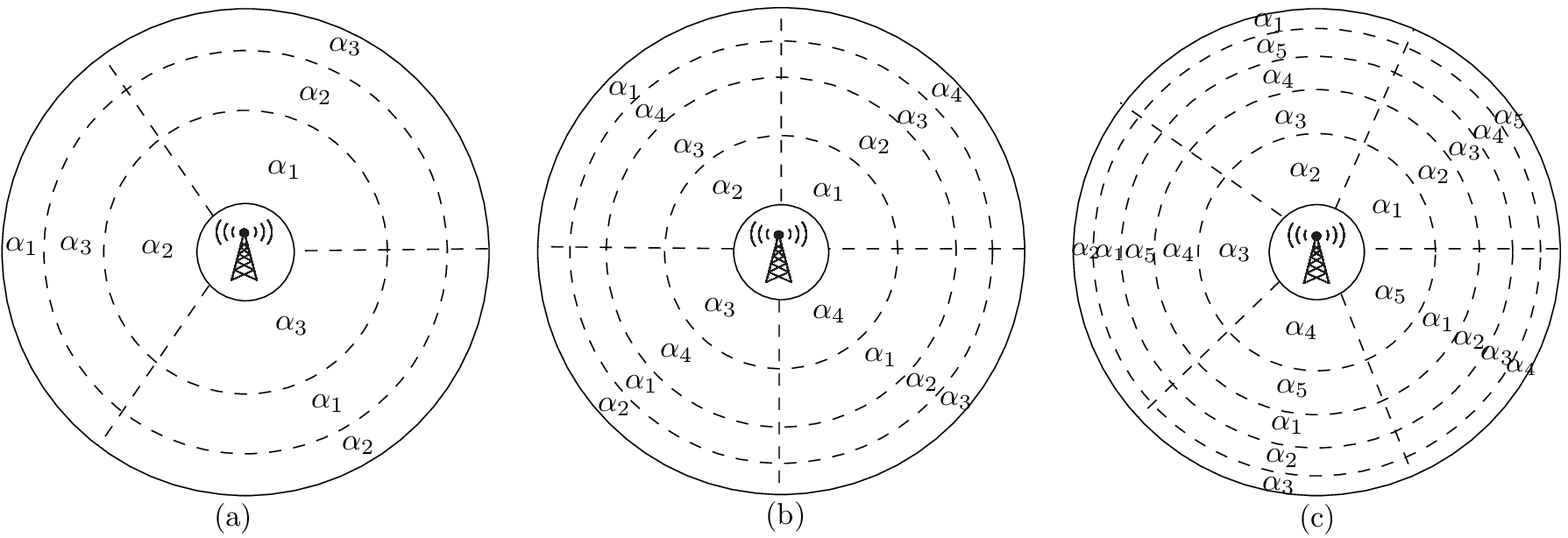}
\caption{Cell planning for the uncoordinated NOMA-HARQ for a) $N = 3$, b) $N = 4$, and c) $N = 5$ active users.}
\label{withPC}
 \vspace{-1.5em}
\end{figure*}
\section{NOMA-HARQ for Grant-Free Access}
In many mMTC scenarios, the BS may not be able to identify the active MTC devices and therefore allocate the optimal power ratios. This becomes very challenging when the number of devices is very large but the number of active devices in each particular time instant is relatively low. In this case, the BS cannot identify and authenticate the active devices at the beginning of each transmission attempt. To solve this problem and in lieu of recent advancements in grant-free multiple access \cite{abbas2020grant}, we propose a grant-free multiple access strategy that can effectively accommodate a large number of devices.  

In the coordinated strategy, the active users first contend for radio resources. Once the users are identified, they will be clustered into several NOMA clusters, in which each cluster has $N_c$ users. The BS then allocates the optimum power ratios obtained for $N_c$ users from (\ref{Opt1}) to the users in each clusters. The users then adjust their transmit powers through the power control algorithm such that the received power at the BS is at the desired level. In the uncoordinated NOMA, the power splitting ratios cannot be optimally assigned to the users, which will lead to performance degradation even though the BS can estimate the number of active users. We assume that the BS can still estimate the number of active users in each time slot by the load estimation strategy proposed in \cite{shirvanimoghaddam2016massive}. The details of load estimation algorithms are beyond the scope of this paper and interested readers are referred to \cite{shirvanimoghaddam2016massive,wiriaatmadja2014hybrid,jang2014spatial} and the references therein for more information.

We propose a dynamic cell planning for the uncoordinated NOMA-HARQ, where each user performs transmit power control according to its segment in the cell. In particular, we assume that the BS performs load estimation and can estimate the number of active users. It then performs cell planning and assigns optimal power splitting ratio to each cell segment. Let $\widehat{N}$ denote the number of active users estimated at the BS. The BS then virtually divides the cell area into $\widehat{N}^2$ regions. In particular, the cell coverage area is partitioned into $\widehat{N}$ annuli and $\widehat{N}$ sectors with the same angles so that the area of all regions are equal. Let $R_o$ denote the cell radius. Then it is easy to show that the radius of the $i^{th}$ ring, denoted by $r_i$, is given by:
\begin{align}
    r_i=\sqrt{\frac{i}{\widehat{N}}}R_o.
\end{align}
Fig. \ref{withPC} shows the cell planning for different numbers of active users. The BS then assigns the optimal power splitting ratios to each segment such that the summation of all power splitting ratios in each annulus and each sector is $1$. In order to ensure that the proposed cell planning does not lead to excessive power usage for users in a particular segment, the BS will rotate the power allocation in each time slot. This makes the average power consumption of users over all cell segments the same. The base station broadcasts the cell planning and power allocation to the users. We assume that the users can determine which segment they are in according to their locations and channel estimation; therefore, they know their power ratios for their transmission. The BS can also determine the configuration of the power levels received at each time slot.

\begin{figure}[t]
\centering
\includegraphics[width=0.6\columnwidth]{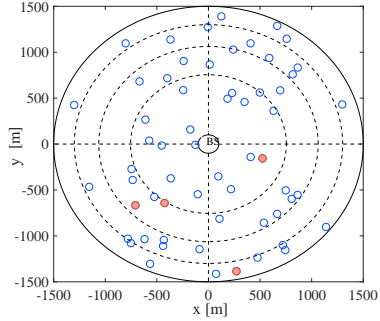}
\caption{Random locations of active users in the proposed dynamic cell planning for $N=4$ users.}
\label{fig:4U}
\end{figure}

It is important to note that in the uncoordinated scenario, more than one user may select the same power level. Fig. \ref{fig:4U} shows a cell partitioned into 16 segments and 4 users are simultaneously active. As can be seen in this figure, two users are located in the same sector and ring; therefore, they will choose the same power level. This will lead to a slight performance degradation. 

\begin{figure*}[t]
\centering
\includegraphics[width=0.66\columnwidth]{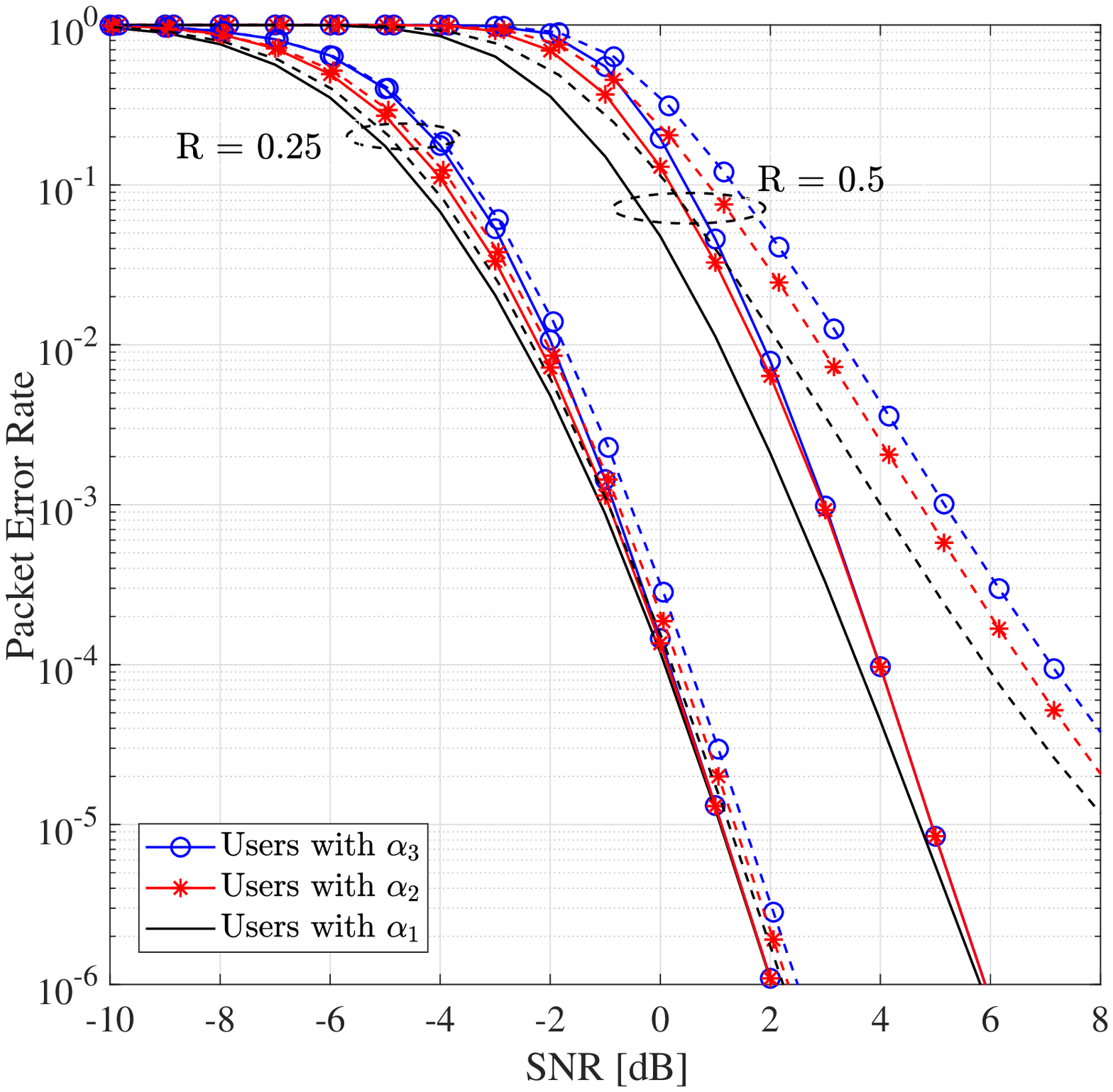}
\includegraphics[width=0.66\columnwidth]{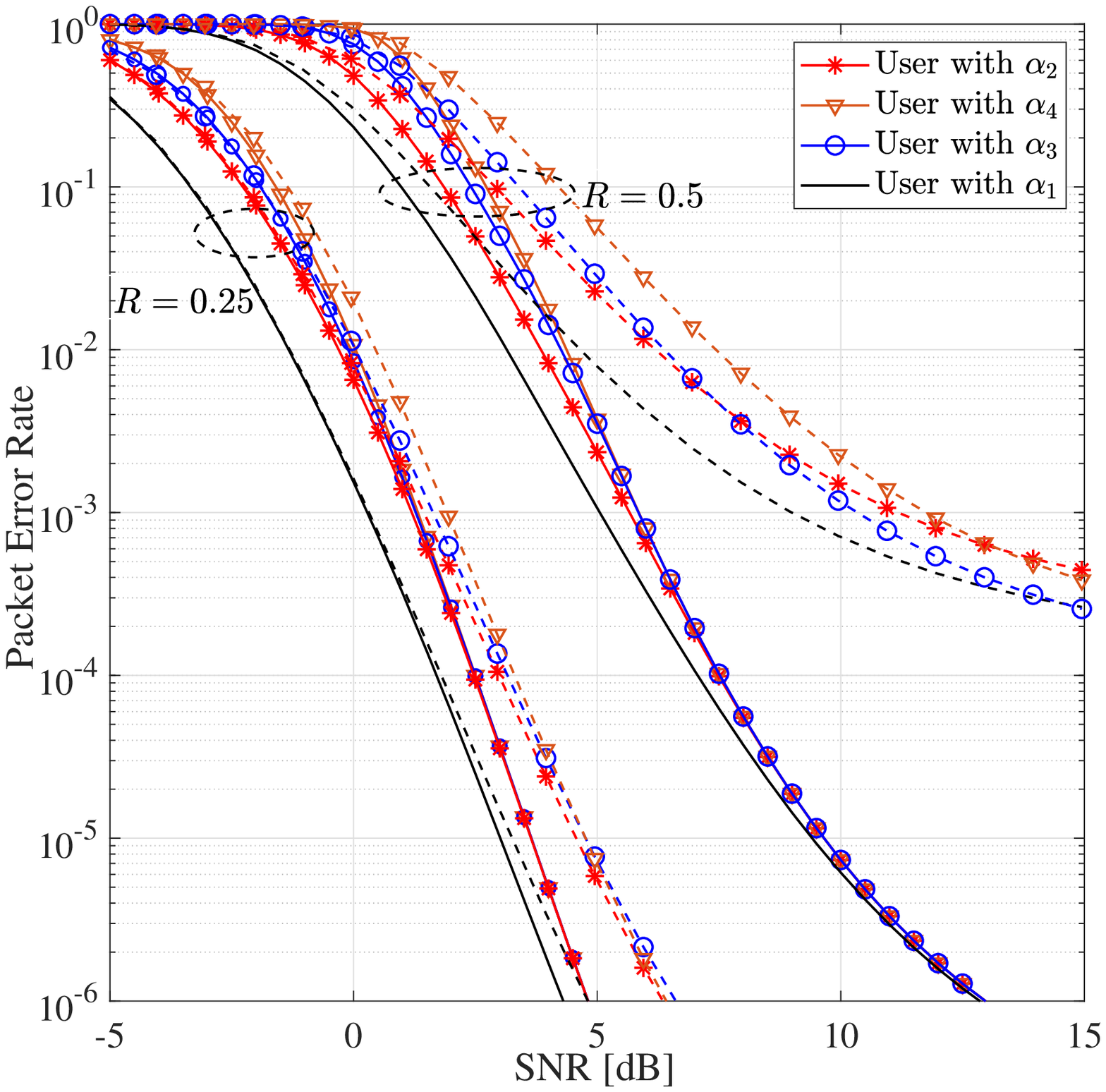}
\includegraphics[width=0.66\columnwidth]{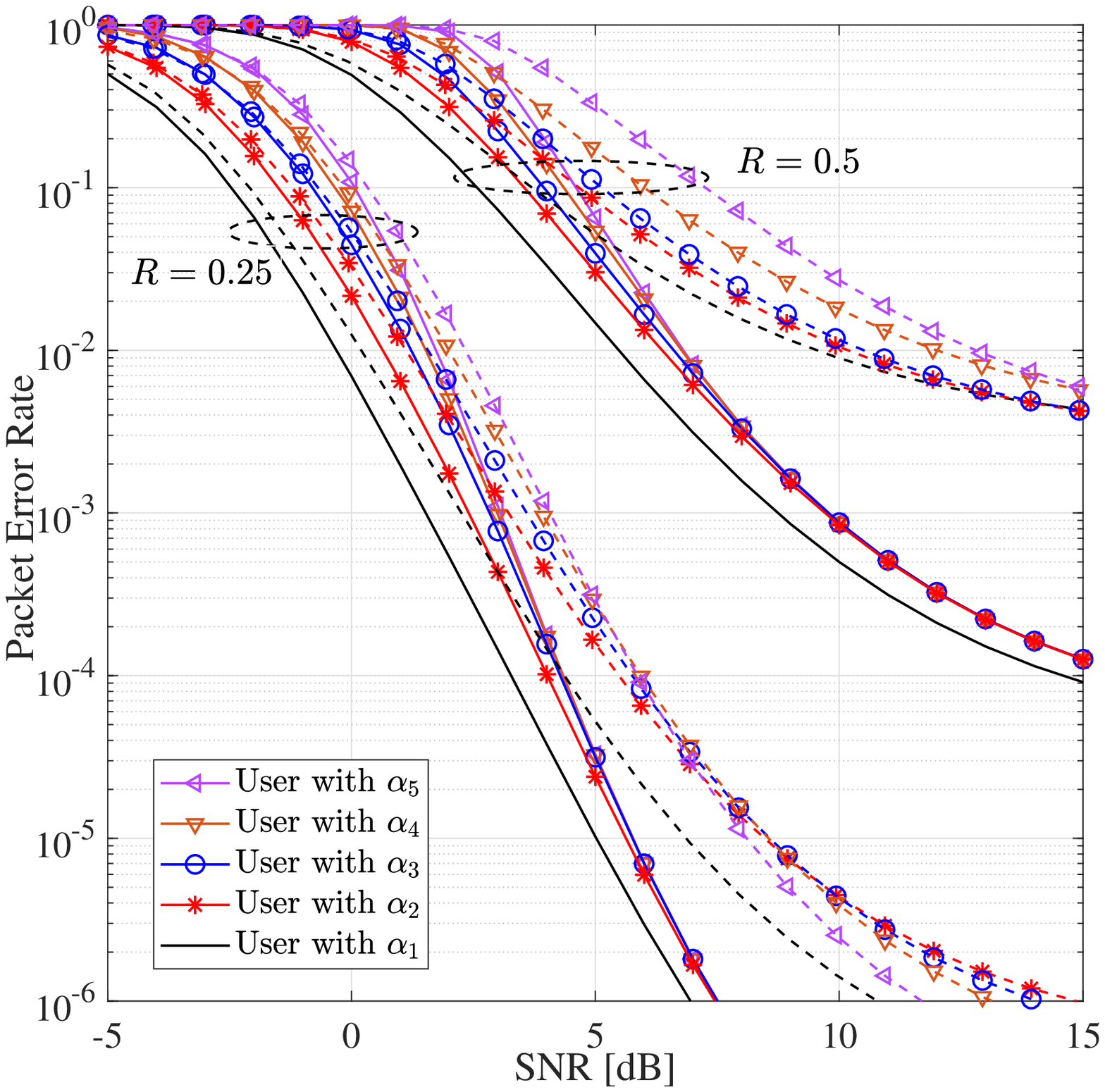}
\\
\footnotesize a) $N=3$ ~~~~~~~~~~~~~~~~~~~~~~~~~~~~~~~~~~~~~~~~~~~~~~b) $N=4$ ~~~~~~~~~~~~~~~~~~~~~~~~~~~~~~~~~~~~~~~~~~~~~~~c) $N=5$
 \vspace{-2ex}
\caption{PER versus SNR performance of 3 users for $n=100$ and two different values of channel code rate, $R=0.25$ and $R=0.5$, where we set the same target packet error rate for all users, $e_t=10^{-2}$.}
\label{fig:PER_C_uC}
 \vspace{-1em}
\end{figure*}
\vspace{-2ex}
\section{Numerical Results}
We assume that users are randomly deployed at fixed locations within a cell with the outer radius of $R_o=1500$ m. In what follows, we show the the packet error rate and throughput performance of NOMA-HARQ  with one retransmission employing Chase combining under the perfect and imperfect load estimation for both coordinated and uncoordinated scenarios with the proposed dynamic cell planning. 

\subsection{Results under Perfect Load Estimation}
\label{sec:perfectload}
We first consider the perfect load estimation, where the BS performs load estimation at the beginning of each time slot. This can be achieved by sending a few training symbols by all active users at the specific power level, so the BS can estimate the number of active users, and then performs cell planning and broadcasts the optimized power for each cell segment to all user. Details of load estimation algorithms are beyond the scope of this paper and interested readers are referred to \cite{shirvanimoghaddam2016massive} and the references therein for further detail. The power splitting ratios for each SNR are obtained from TABLE II for the target PER of $10^{-2}$. 

\begin{figure*}[t]
\centering
\includegraphics[width=0.61\columnwidth]{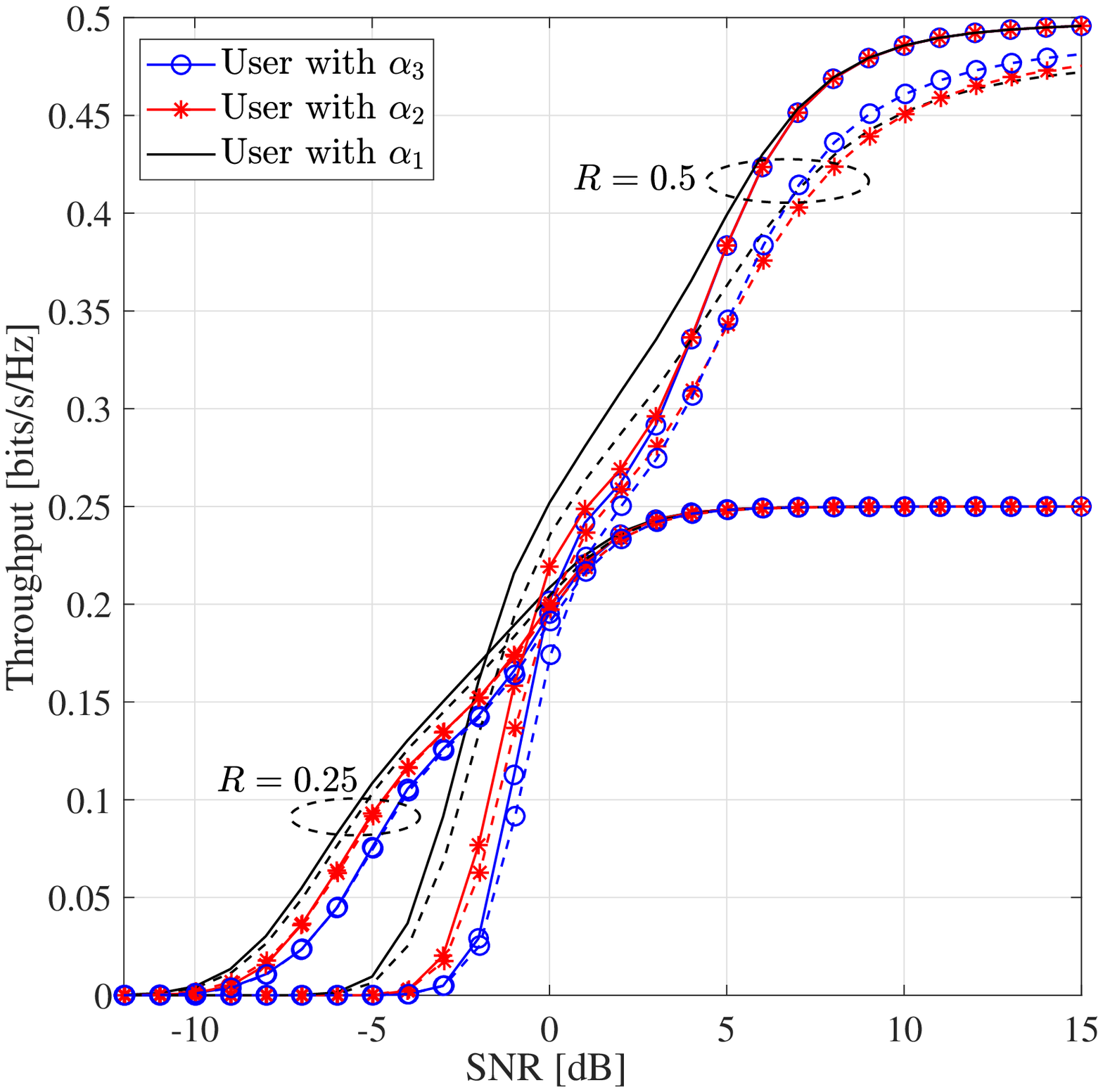}
\includegraphics[width=0.61\columnwidth]{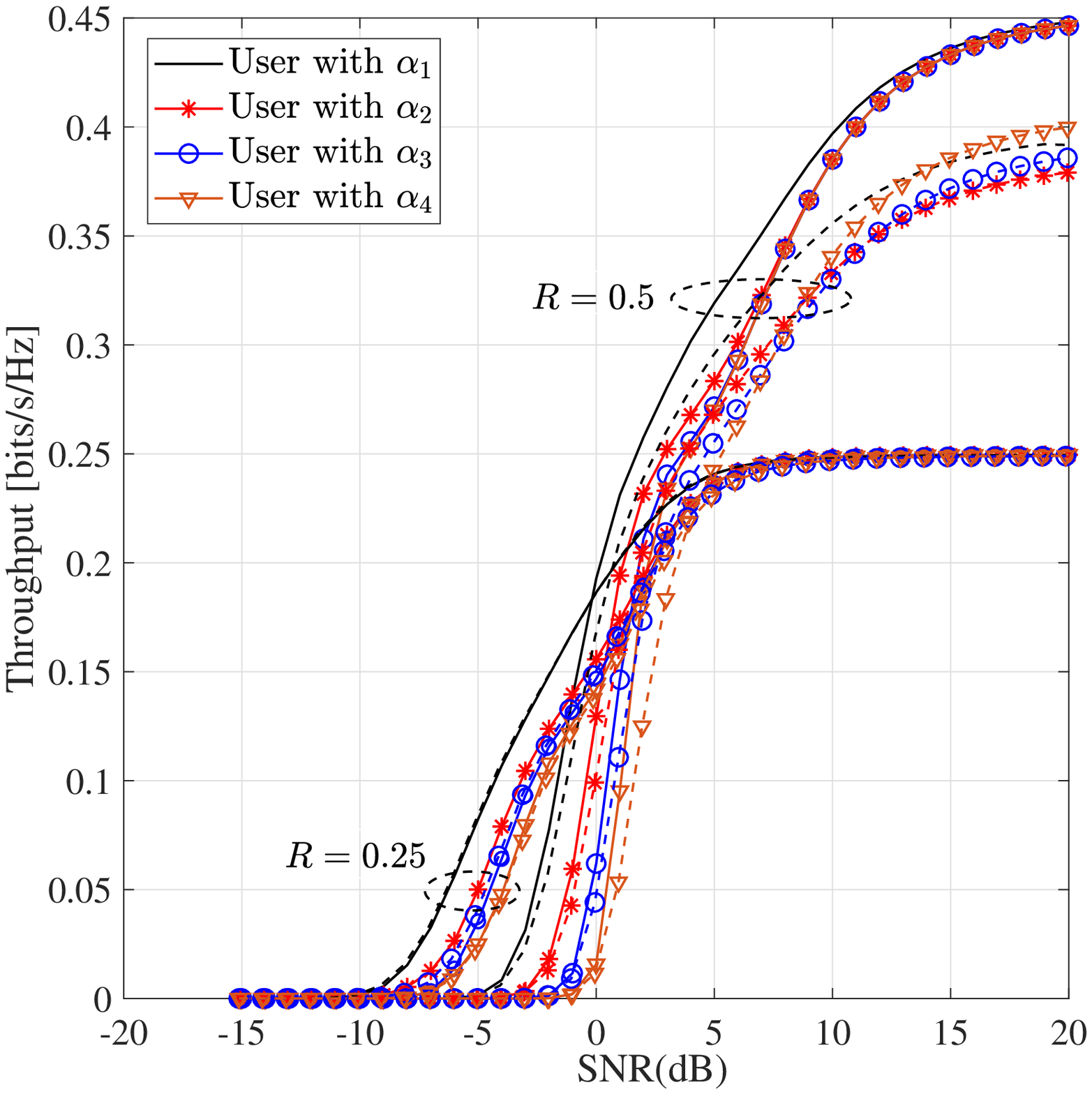}
\includegraphics[width=0.61\columnwidth]{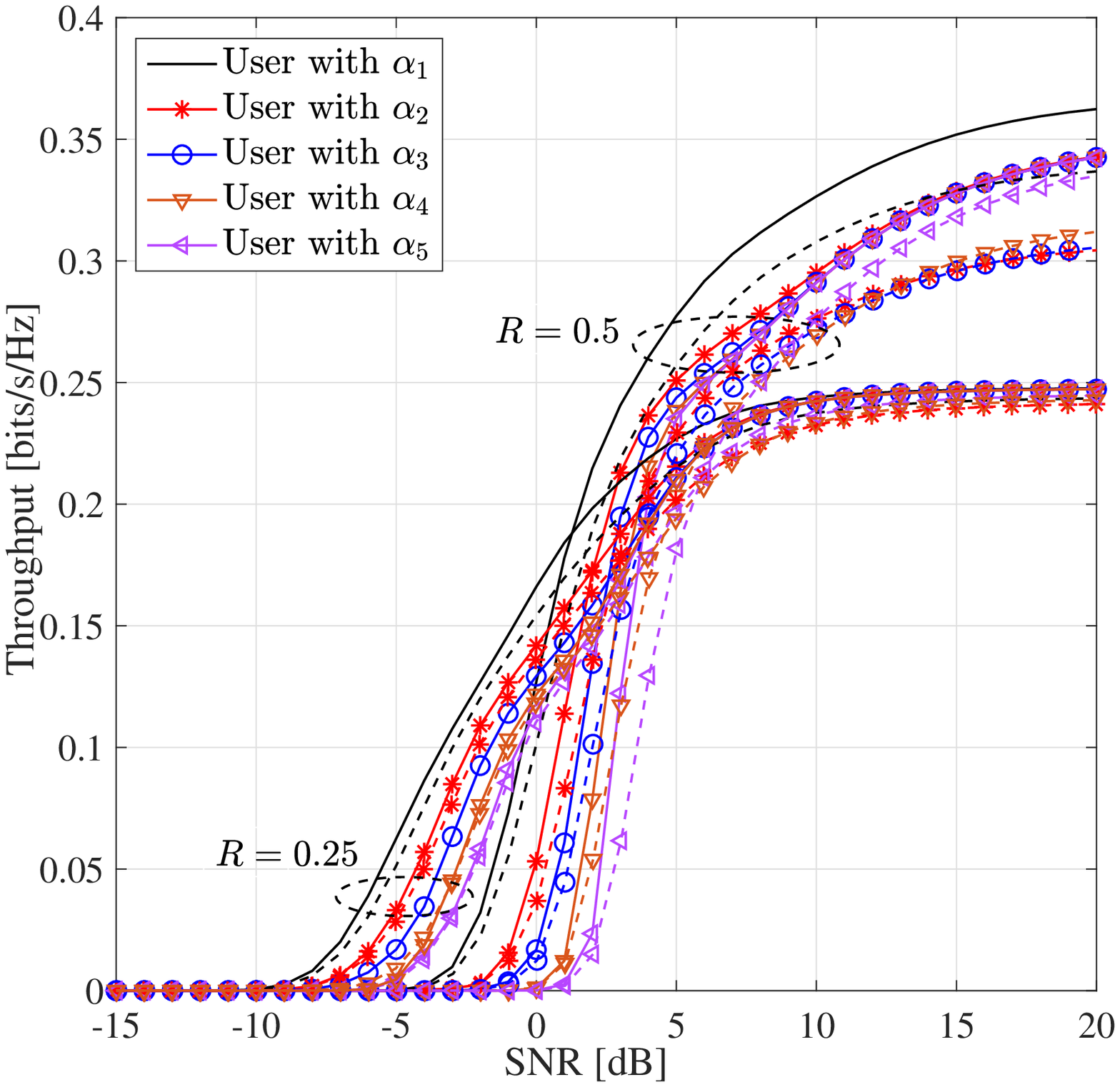}\\
\footnotesize a) $N=3$ ~~~~~~~~~~~~~~~~~~~~~~~~~~~~~~~~~~~~~~~~~~~~~~b) $N=4$ ~~~~~~~~~~~~~~~~~~~~~~~~~~~~~~~~~~~~~~~~~~~~~~~c) $N=5$
 \vspace{-2ex}
\caption{Throughput versus SNR for NOMA-HARQ when $n=100$ with two different code rates, $R=0.25$ and $R=0.5$. The target packet error rate for all users is $e_t=10^{-2}$. Solid and dashed lines show the result for the coordinated and uncoordinated NOMA-HARQ with one retransmission, respectively.}
\label{fig:Th_C_U}
 \vspace{-2em}
\end{figure*}
\begin{figure}[t]
\centering
\includegraphics[width=0.8\columnwidth]{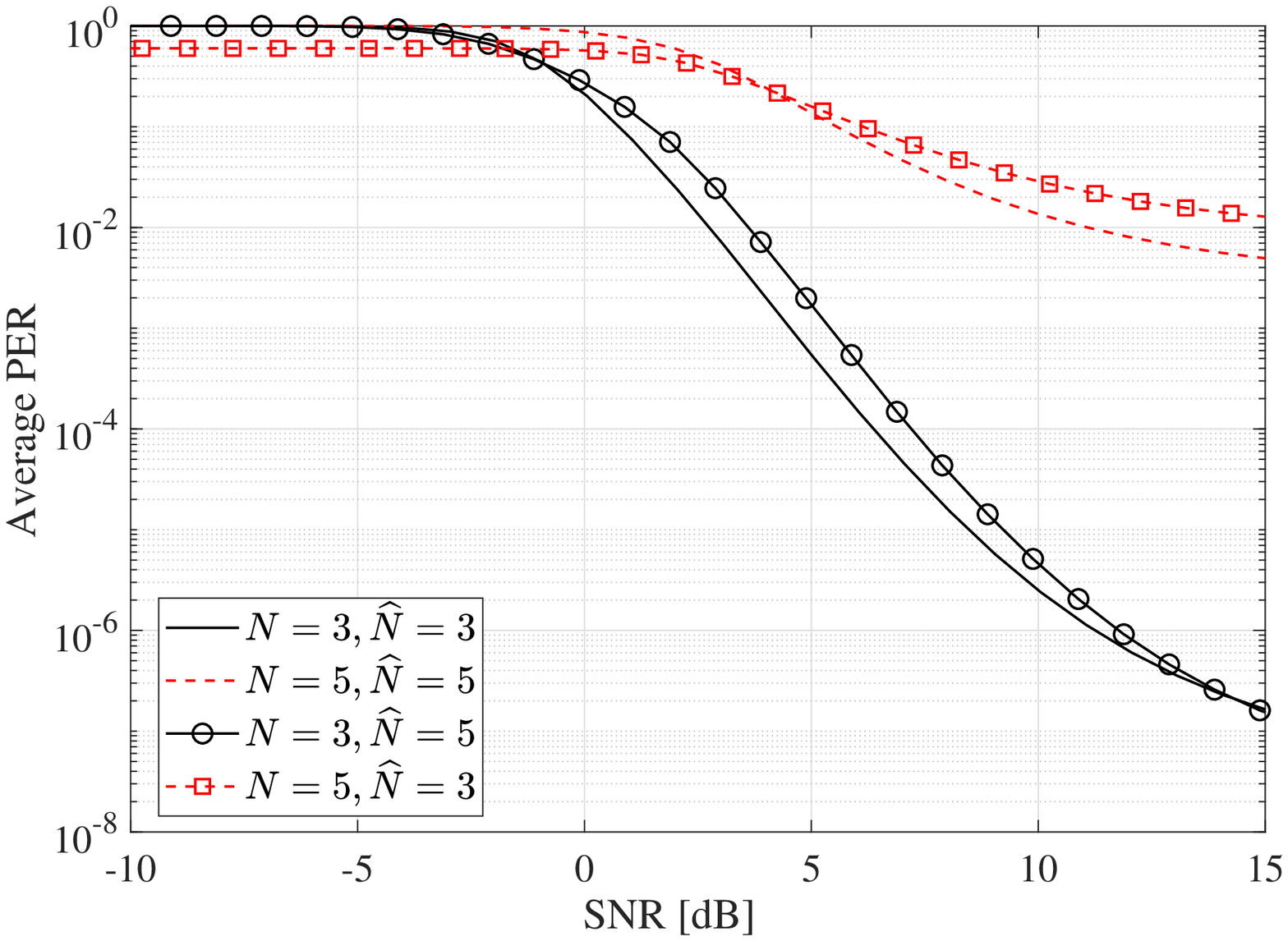}
\caption{Average PER versus SNR for the uncoordinated NOMA-HARQ with one retransmission when $R=0.5$ and $n=100$.}
\label{ac_inac}
\end{figure}
Fig. \ref{fig:PER_C_uC} shows the PER versus the SNR for NOMA-HARQ with one retransmission at different channel loads when the power splitting ratios are obtained at the target error probability of $10^{-2}$. As can be seen, when the number of active users and the code rate are low, the uncoordinated scenario performs very close to the coordinated one. However, when the channel load or the code rate increase, the uncoordinated NOMA-HARQ will experience worse performance compared with the coordinated scenario. This is mainly due to the fact that when the channel load increases, the probability that more than one user is active in each cell segment in each time slot increases, which leads to a degradation of the PER performance. In fact, in the coordinated scenario the active users are always transmitting with the optimal power splitting ratios as the BS can identify and allocate each user with the optimal power ratio. However, in the uncoordinated scenario the users are choosing their transmit power according to their location and cell segment; therefore, they are not necessarily transmitting with the optimal configuration of power ratios. However, the results shown in Fig. \ref{fig:PER_C_uC} indicates that the proposed cell planning is effective in order to support a large number of devices in an uncoordinated manner. One can choose a sufficiently low channel code rate to guarantee the performance and minimize the loss in comparison with the coordinated scenario. 

Fig. \ref{fig:Th_C_U} shows the throughput performance for both the coordinated and uncoordinated scenarios at the target PER of $10^{-2}$. As can be seen, the uncoordinated NOAM-HARQ with the proposed dynamic cell planning achieve almost the same throughput as the coordinated scenario. The slight decrease in the throughput in the high SNRs is mainly due to the fact that the users in the uncoordinated scenario do not necessarily transmit with the optimal configuration of the power ratios. When a lower code rate is chosen, the gap will diminish in high SNRs. Therefore, one needs to choose a lower code rate to properly support a large number of active devices and guarantee the desired level of reliability and throughput. It is important to note that OMA-HARQ will perform poorly in the uncoordinated scenario as the devices will randomly transmit and the performance will be significantly degraded due to collision between packets. One may consider some backoff strategies or class-barring techniques \cite{shirvanimoghaddam2017massive} to improve the performance of OMA; however, these will lead to poor throughput performance.  
\vspace{-2ex}
\subsection{Results under Imperfect Load Estimation}
As mentioned earlier, the load estimation at the BS may not be accurate. The load estimation also adds extra overhead as devices needs to send pilots at the beginning of each time slot, so the BS can perform load estimation. This significantly reduces the throughput and increase the latency. Therefore, such approaches may not be suitable for delay sensitive applications. In the absence of accurate load estimation, the BS however can obtain statistical channel load profile from the output of the receiver in previous time slots. In this case, the BS will perform cell planning based on the inaccurate load estimation. Let $\widehat{N}$ denote the estimated load, then the BS divides the cell into $\widehat{N}^2$ segments and broadcasts the information regarding the optimized power ratios for each segment to all users. The number of active users in each time slot, denoted by $N$, may be larger than, equal to, or less than $\widehat{N}$. In this part, we are interested in cases where $N>\widehat{N}$ or $N<\widehat{N}$. The case where $N=\widehat{N}$ was previously discussed in Section \ref{sec:perfectload}.

Fig.~\ref{ac_inac} shows the average packet error rate of active users versus the average received SNR at the BS for different numbers of active users and cell segments. The power splitting ratios for each SNR are obtained from TABLE II for the target PER of $10^{-2}$. As can be seen, when the cell is partitioned into 9 segments, i.e., $\widehat{N}=3$, and the number of active users is $N=5$, the average PER is very close to the optimal case when  $N=\widehat{N}=5$. In this case, more than one user have always selected the same power ratio and the overall received power at the BS may exceeds $P_0$. This might be an indicator for the BS to slightly adjust its load estimation. It is important to note that when more than one user select the same power level, the BS may fail decoding both users. This however can be resolved by using powerful channel codes with sufficiently low code rates.

When the estimated load is $\widehat{N}=5$ and the number of active users $N=3$, the average PER is close to the optimal case when $N=\widehat{N}=3$. This is because, the users are now sending with smaller power levels, and the total received power at the BS is less than $P_0$. The BS, then needs to adjust the load estimation or increase the reference power $P_0$. The results shown in Fig. \ref{ac_inac} indicates that the proposed scheme is not very sensitive to the load estimation algorithm and can achieve the desired level of PER for all active users. The BS may however prefer overestimation rather than underestimation of the load, in order to achieve a higher reliability. This is very important for multi-channel grant-free NOMA, where each user randomly chooses a channel (or a subband) for its transmission and chooses the power level according to its location in the cell. The BS may also consider different cell planning over each channel.

It is important to note that the BS needs to determine the power levels selected by active users when performing the SIC. This can be achieved by first estimating the number of active users and then estimating the selected powers. As the power levels for each $\widehat{N}$ and the desired level of reliability are unique, the BS can easily find the number of active users and selected power levels by testing all possible cases of choosing a set of power levels with repetition. One can also design a more sophisticated estimation technique which is beyond the scope of this paper. 
\vspace{-2ex}
\section{Conclusion}
We considered an uplink NOMA-HARQ system, where multiple active users share the same radio resources and are allowed to retransmit their packets only once to minimize the latency. We analyzed the dynamics of the NOMA-HARQ with one retransmission using a Markov model and the packet error rate and throughput were accordingly characterized. We then formulated and numerically solved two different optimization problems, 1) in a power constrained scenario and 2) in a reliability constrained scenario, to find the optimal power splitting ratios. We further proposed a dynamic cell planning for the uncoordinated transmission, where the users choose their power levels according to their location in the cell. Numerical results show that both coordinated and uncoordinated NOMA-HARQ with a limited number of retransmissions can achieve the desired level of reliability with the guaranteed latency using a proper power control strategy. Results also show that NOMA-HARQ achieves a higher throughput compared to the orthogonal multiple access scheme with HARQ under the same average received power constraint at the base station. 
\vspace{-2ex}
\appendices
\section{Proof of Lemma 1}
As we allow a maximum of one retransmission, a user at state $\mathrm{F}$ or $\mathrm{S}$ will not directly transit to state $\mathrm{F}$. Therefore, $\pi_{\mathbf{J}\rightarrow \mathbf{J}'}=0$ when the transition indicates that there is a user $i$ that transits from $J_i\in\{\mathrm{F,S}\}$ to $J'_i=\mathrm{F}$. This proves the first line of (\ref{eq:transprob}).

Since $I_i$ denotes the user with the highest SINR at the $i^{th}$ stage of SIC, it is clear that all the remaining users will either fail or transit to the state $\mathrm{R}$, if user $I_i$ transit to state $\mathrm{F}$ or $\mathrm{R}$. This is because the SIC only proceeds to the next stage when the decoding succeeds in the current stage. Therefore, if the transition from state $\mathbf{J}$ to $\mathbf{J}'$ indicates that the user $I_j$ goes to state $\mathrm{S}$ when there is a user $I_i$ with $i<j$ that transits to state $\mathrm{F}$ or $\mathrm{R}$, then $\pi_{\mathbf{J}\rightarrow \mathbf{J}'}=0$.  This proves the second line of (\ref{eq:transprob}).

If the system transits from state $J$ to state $J'$ and none of the above mentioned conditions hold, the BS will decode the users using SIC according to the decoding order in each stage, i.e., $I_i$. That is the BS first decode user $I_1$, and if successful then moves to decode user $I_2$, and so forth. The SIC stops whenever the decoding failed, say at stage $m$. All the remaining users will transit to either state $\mathrm{F}$ or $\mathrm{R}$ depending whether they have already performed the retransmission or not. All users from $I_1$ to $I_{m-1}$ are transiting to state $\mathrm{S}$, while user $I_{m}$ will transit to state $\mathrm{F}$ or $\mathrm{R}$. For the user transiting from state $\mathrm{F}$ or $\mathrm{S}$ to state $\mathrm{S}$, the BS has only one copy of its packet; therefore, the transition probability for user $I_i$ will be given by $q_{I_i}=1- \epsilon\left(\gamma^{(i)}_{I_i}\right)$, where $\gamma^{(i)}_{I_i}$ is given by (\ref{eq:SINR-SIC-J}). For the user transiting from state $\mathrm{F}$ or $\mathrm{S}$ to state $\mathrm{R}$, the transition probability for user $I_i$ will be given by $q_{I_i}=\epsilon\left(\gamma^{(i)}_{I_i}\right)$. For the user transiting from state $\mathrm{R}$ to state $\mathrm{S}$, the BS has two copies of its packet; therefore, the transition probability for user $I_i$ will be given by $q_{I_i}=1- \epsilon\left(\gamma^{(i)}_{I_i}\right)$, where $\gamma^{(i)}_{I_i}$ is given by the second line of (\ref{eq:SINR-SIC-J}). For the user transiting from state $\mathrm{R}$ to state $\mathrm{F}$, the transition probability for user $I_i$ will be given by $q_{I_i}= \epsilon\left(\gamma^{(i)}_{I_i}\right)$. The probability of transiting from state $\mathbf{J}$ to state $\mathbf{J}'$ can be easily calculated by multiplying the state transition probability for users $I_1$ to $I_m$. 
\vspace{-2ex}
\section{Proof of Lemma 2}
The packet of the $i^{th}$ user is declared as failed in one of the following two cases, 1) when the $i^{th}$ user is at state $\mathrm{F}$ and 2) when the $i^{th}$ user is at state $\mathrm{R}$ and transits to state $\mathrm{F}$. Let $\mathcal{F}_{i}$ denote the set of all the states that the $i^{th}$ user is at state $\mathrm{F}$, then the probability that case 1 occurs is given by $\sum_{w\in {{\mathcal{F}}_{i}}}{{{p}_{w}}}$. Let $\mathcal{R}_{i}$ denote the set of all the states that the $i^{th}$ user's state is $\mathrm{R}$, then the probability that it transits to state $\mathrm{F}$, is given by $\pi_{w,j}$, where $w\in 
\mathcal{R}_i$ and $j\in\mathcal{F}_i$. By taking the sum over all $w$ and $j$, the probability that case 2 occurs is given by $\sum_{w\in 
\mathcal{R}_i}p_w\left(\sum_{j\in\mathcal{F}_i}\pi_{w,j}\right)$. This completes the proof.
\vspace{-2ex}
\section{Proof of Lemma 3}
When a user is at either state $\mathrm{F}$ or state $\mathrm{S}$ and
transits to state $\mathrm{S}$, only one transmission is required to deliver the packet. The probability
that a packet will be successfully decoded is given by (\ref{eqpsuccess}).
Each information packet of user $i$ will be decoded with only
one transmission with probability $p_s^{(i)}$; otherwise user $i$ needs
to re-transmit the packet which will happen with probability $1-p_s^{(i)}$. The number of packets that user $i$ needs to send to potentially deliver each information packet is a Bernoulli random variable with success probability $p_s^{(i)}$, where the success and fail events respectively correspond to 1 packet and 2 packets delay. The total number of packets to be sent to deliver $M$ information packets is then a Binomial random variable with the distribution given in (\ref{eqbinomial}).
  \ifCLASSOPTIONcaptionsoff
  \newpage
  \fi
\bibliographystyle{IEEEtran}
\bibliography{IEEEabrv,Reference}

\vspace{-4em}
\begin{IEEEbiography}[{\includegraphics[width=1in,height=1.25in,clip,keepaspectratio]{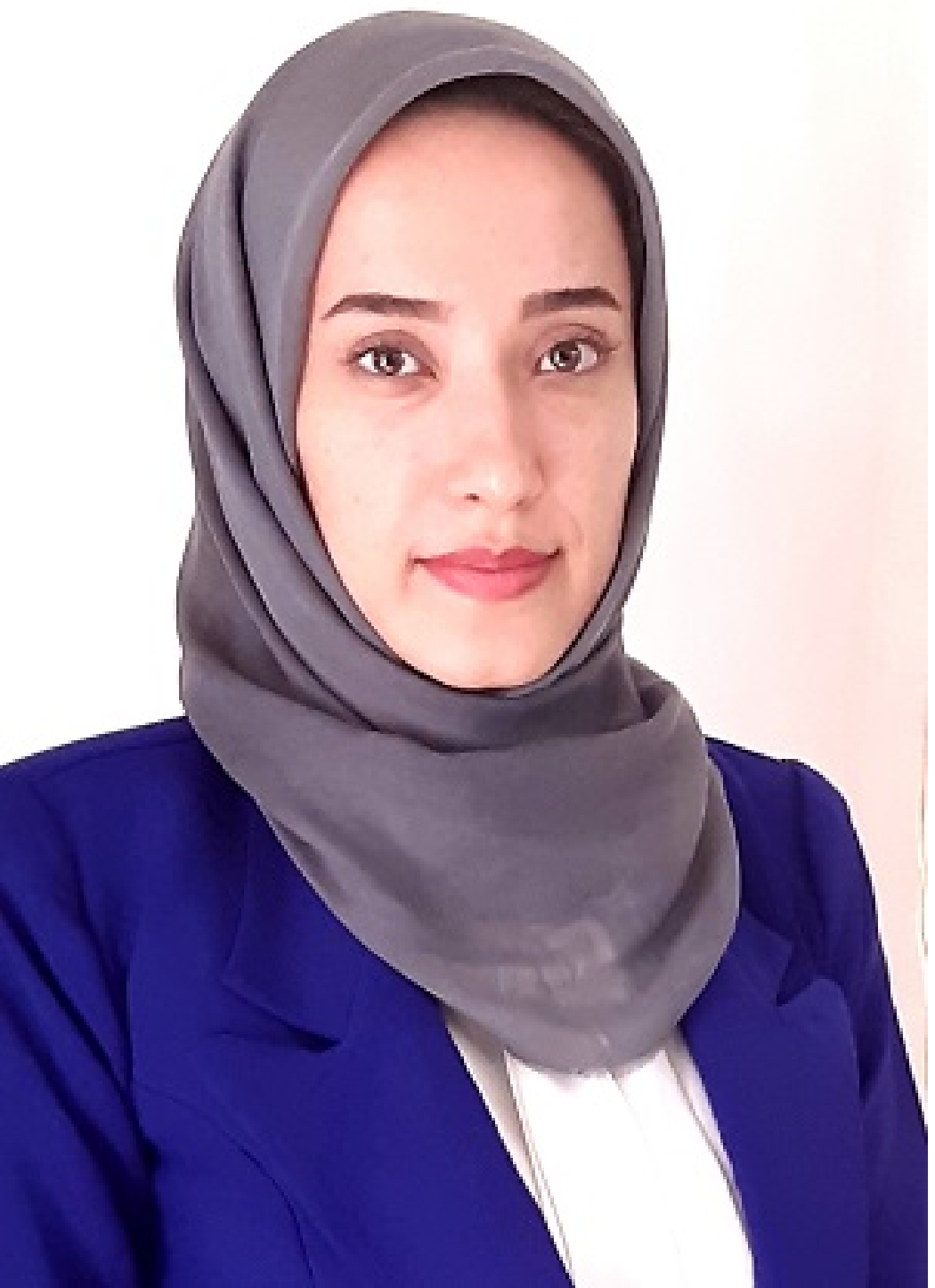}}]{Fatemeh Ghanami} 
 received the M.Sc. degree (Hons.) in Communications Engineering from Ferdowsi University of Mashhad, Mashhad, Iran, in 2014, where she is currently pursuing the Ph.D. degree in communication systems with the Department of Electrical Engineering. Her current research interests include information theory, wireless communication, physical layer security, Internet of Things, URLLC communications, and machine learning.
\end{IEEEbiography}
\vspace{-5em}
\begin{IEEEbiography}[{\includegraphics[width=1in,height=1.25in,clip,keepaspectratio]{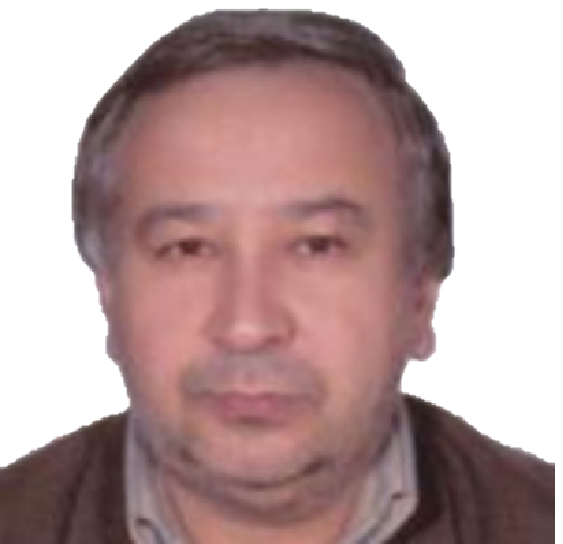}}]{Ghosheh Abed Hodtani}  received his B.Sc. degree in Electronics Engineering and M.Sc. degree in Communications Engineering from Isfahan University of Technology, Isfahan, Iran, in 1985 and 1987, respectively, and joined the Department of Electrical Engineering with Ferdowsi University of Mashhad, Mashhad, Iran, in 1987. He decided to pursue his studies in 2005 and received the Ph.D. degree from Sharif University of Technology, in 2008; and was promoted to a full professor in 2016. Prof. Hodtani is the author of a textbook on Electrical Circuits, and his research interests are in multi-user information theory, communication theory, wireless communications, and signal processing. He is recipient of the best paper award at the IEEE International Conference on Telecommunications in 2010. He is a member of the Technical Program and Steering Committees of the Iran Workshop on Communication and Information Theory.
\end{IEEEbiography}
\vspace{-5em}
\begin{IEEEbiography}[{\includegraphics[width=1in,height=1.25in,clip,keepaspectratio]{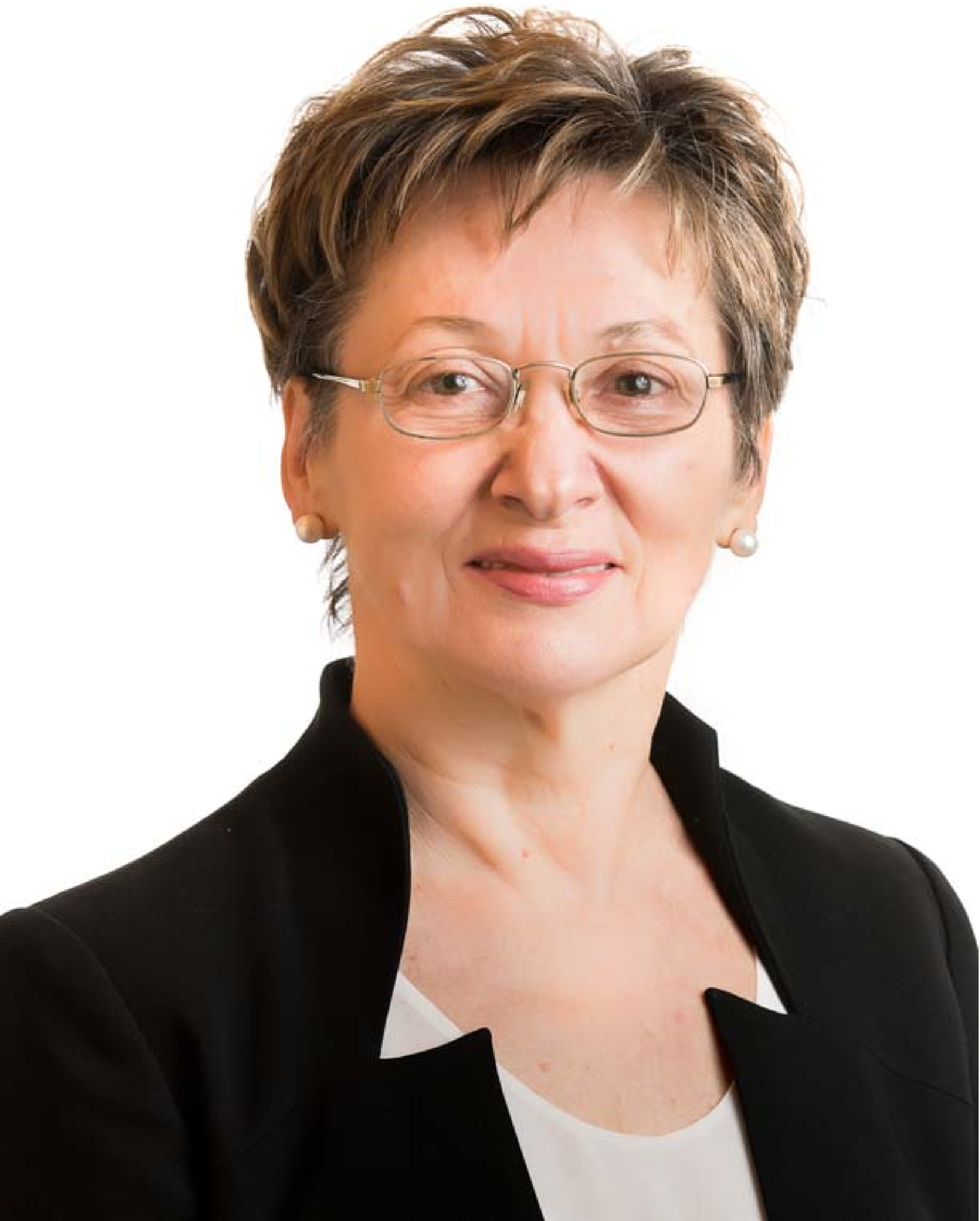}}]{Branka Vucetic} (F'03) received the B.S.E.E.,
M.S.E.E., and Ph.D. degrees in electrical engineering from the University of Belgrade, Belgrade, Serbia, in 1972, 1978, and 1982, respectively. She is an ARC Laureate Fellow and the Director of the Centre of Excellence for Internet of Things
and Telecommunications, University of Sydney, Sydney, NSW, Australia. Her current research interests include wireless networks and Internet of Things (IoT). In the area of wireless networks, she explores ultra-reliable, low-latency techniques and
transmission in millimeter wave frequency bands. In the area of the IoT, Vucetic works on providing wireless connectivity for mission critical applications. Ms. Vucetic was a Fellow of the Australian Academy of Science and the Australian Academy of Technological Sciences and Engineering.
\end{IEEEbiography}
\vspace{-5em}
\begin{IEEEbiography}[{\includegraphics[width=1in,height=1.25in,clip,keepaspectratio]{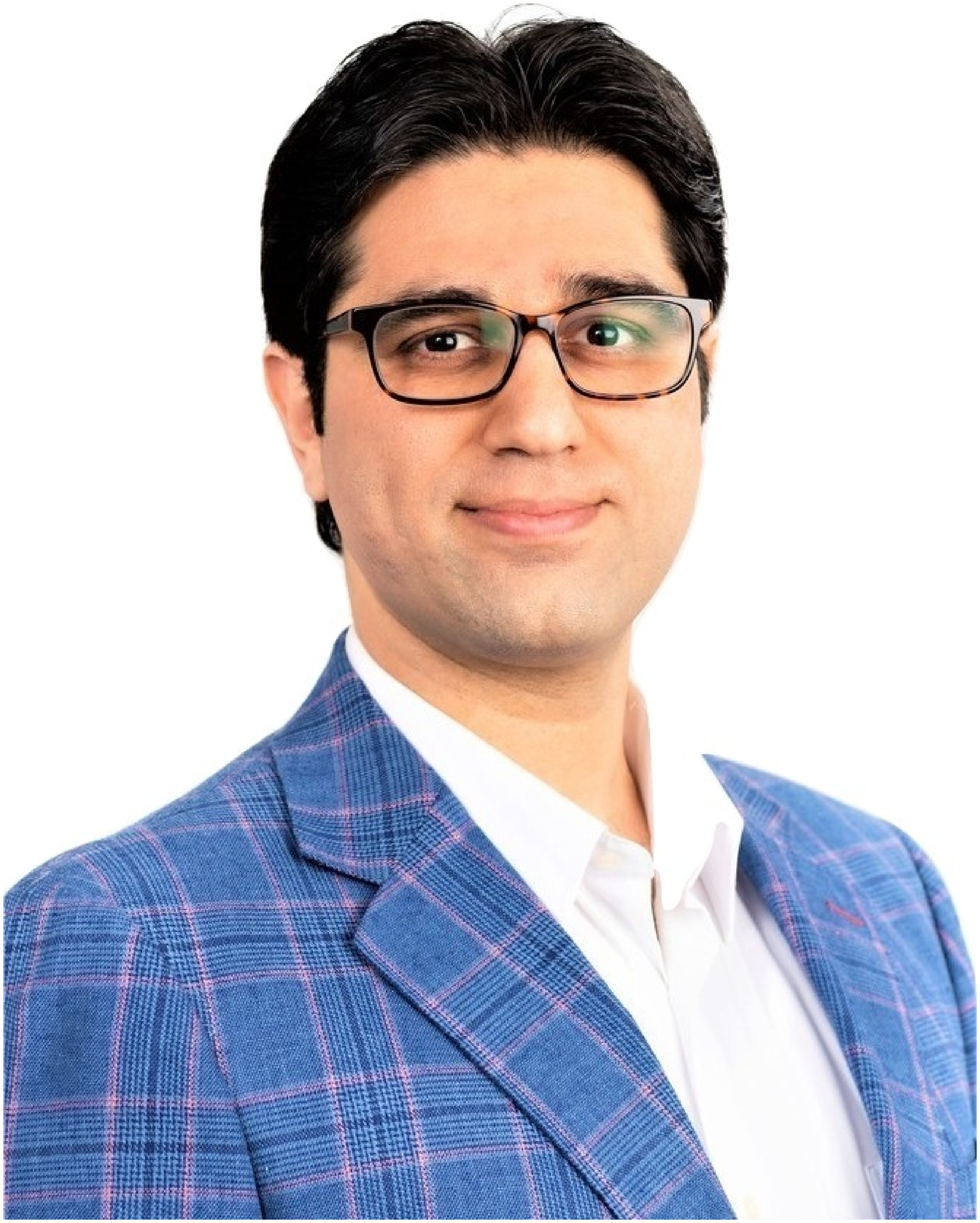}}]{Mahyar Shirvanimoghaddam} (S'19)
 received his BSc and MSc degrees, both in Electrical Engineering, with 1st Class Honor in 2008 and 2010, respectively from Sharif University of Technology and University of Tehran, Iran. He received his PhD in Electrical Engineering from The University of Sydney, Australia, in 2015. He is currently a Lecturer at Centre for IoT and Telecommunications, The University of Sydney. His general research interests include Coding and Information Theory and Internet of Things technologies. He is an IEEE Senior Member and a Fellow of the Higher Education Academy (FHEA). In 2018, he was selected as one of the Top 50 Young Scientists in the World by the World Economic Forum for his contributions to the development of IoT technologies.
\end{IEEEbiography}

\end{document}